\documentclass[12pt]{article}

\usepackage{rotate}
\usepackage{epsf}
\usepackage{psfig}

\def\pt{$ p_{t}$ } 
\def\xf{$ x_F$ } 
 
\def\u{$\Upsilon$ } 
\def\JP{$\psi'$ } 
\def\J{$J/\psi$ }
\def\up{$\Upsilon^{\prime}$ } 
\def\upp{$\Upsilon^{\prime\prime}$ }

\begin{document} 

\epsfclipon
 
\title{\bf HIGH-ENERGY HADRON-INDUCED DILEPTON PRODUCTION FROM NUCLEONS 
AND NUCLEI}
\author{{\it P.L. McGaughey, J.M. Moss, and J.C. Peng}}
\maketitle

\begin{center}
Los Alamos National Laboratory
\end{center}

\bigskip

\noindent{CONTENTS}

\noindent{1. INTRODUCTION}

\noindent{2. THE DRELL-YAN PROCESS}

2.1 Drell-Yan Process in the Parton Model

2.2 QCD and the Drell-Yan Process

2.3 Factorization and Nuclear Size Effects

2.4 Experimental Detection of Lepton Pairs

2.5 Backgrounds in Inclusive Drell-Yan Experiments

\noindent{3. DRELL-YAN PROCESS AND PARTON DISTRIBUTIONS}

3.1 $\bar d/\bar u(x)$ Asymmetry of the Proton

3.2 Polarized Drell-Yan and $W^\pm$ Production

3.3 Charge Symmetry Violation of Parton Distributions

3.4 Parton Distributions of Mesons and Hyperons

3.5 Nuclear Dependence of the Drell-Yan Process

3.6 Shadowing

3.7 Multiple Scattering and Energy Loss

\noindent{4. QUARKONIUM PRODUCTION}

4.1 Quarkonium Production in Hadronic Collisions

4.2 Nuclear Dependence of Quarkonium Production

4.3 Nuclear Dependence of Open Charm Production

4.3 Nuclear Effects in Quarkonium Production and the Quark-Gluon Plasma

\noindent{5. FUTURE PROSPECTS}

\vskip 0.5in

\begin{abstract}
We review the production of high-mass lepton pairs in fixed-target 
experiments, including both Drell-Yan (DY) and heavy 
quarkonium production ($J/\psi$, $\psi '$, $\Upsilon(1S)$, 
$\Upsilon(2S)$, and $\Upsilon(3S)$). In recent 
years DY data has become standard input to the determination of 
parton density distributions. DY data has recently yielded the
first measurement of the 
$x$-dependence of the $\bar d/\bar u$ asymmetry of the proton.
Similar to the observations in deeply inelastic scattering, 
precision measurements of the nuclear dependence of the proton-induced 
DY process exhibit shadowing at small target momentum fraction, $x_2$. 
There is, however, no evidence of
enhanced DY production from nuclear targets. Mean transverse 
momenta of DY pairs is observed to increase with target mass. These 
data, analyzed within a new theoretical framework, provides an estimation
of the energy loss of fast quarks in nuclear matter. In contrast 
to the DY process, there are large nuclear effects in the production of 
all quarkonia. These effects show strong dependence on \pt and 
$x_F$, but do not scale with $x_2$.
Statistically less precise measurements of $D$ meson (open-charm) production 
at small $x_F$ are consistent with no nuclear dependence.
\end{abstract}

\newpage
\noindent{1. INTRODUCTION}

\begin{figure}                                                  
\centerline{
\psfig{figure=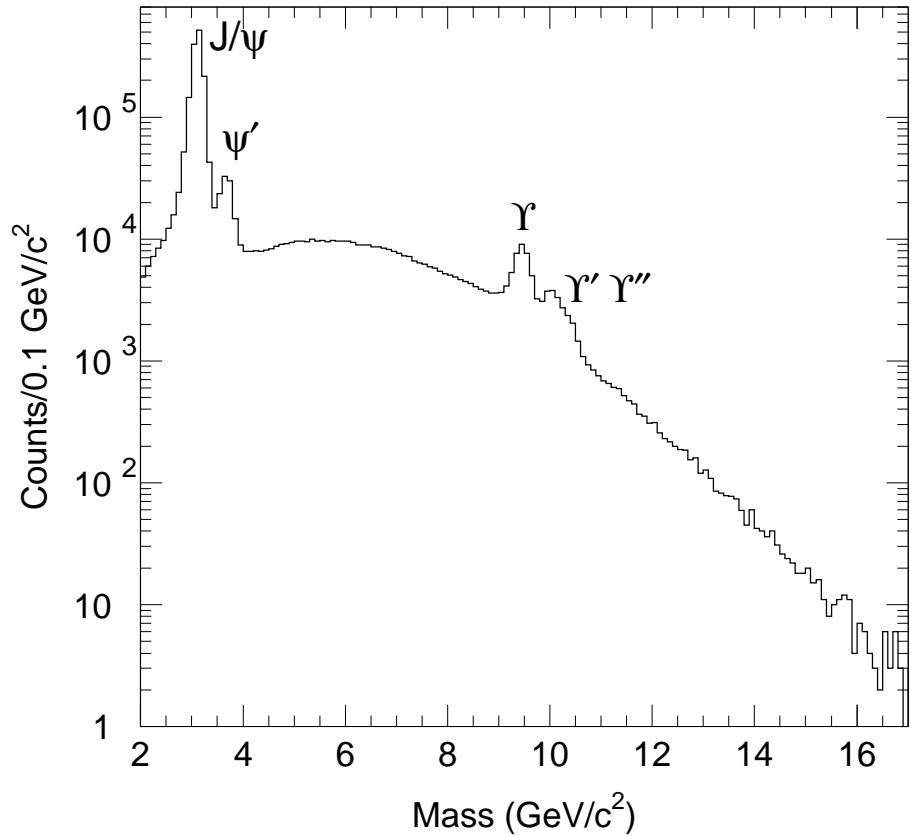,height=4.38in}}
\caption{Combined dimuon mass spectrum from Fermilab E866~\cite{e866}:
$p + p$ and $p + d$ collisions 
at 800 GeV/c. The shape of the continuum 
results from the mass dependence of the DY process folded with the 
acceptance of the spectrometer.}
\label{fige866a} 
\end{figure}

\begin{figure}    
\centerline{
\psfig{figure=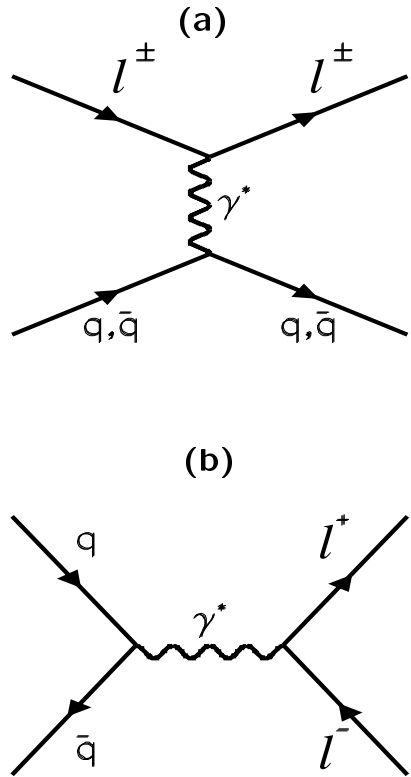,height=4.38in}}
\caption{Feynman graphs for the electromagnetic processes (a)
deeply inelastic lepton scattering and (b) the Drell-Yan process.}
\label{disdy}
\end{figure}

The experimental detection of high-mass\footnote{In the context of this 
review, high mass is $M_{l^+l^-}\geq M_{J/\psi}$} lepton pairs produced 
in hadronic reactions
has a long and rich history. The famous quarkonium 
states that revealed the existence of the charm and beauty quarks in 
the 1970s were discovered through their dilepton decay branches. They are
superimposed on a continuum, which was anticipated theoretically in 
1970~\cite{drell}, and is now known as the Drell-Yan (DY) 
process (Figure~\ref{fige866a}). The DY process, electromagnetic 
quark-antiquark annihilation, is shown diagramatically in 
Figure~\ref{disdy}, along with its close relative, deeply inelastic 
lepton scattering (DIS). By 1980, DY production 
was already a source of information about antiquark structure of the nucleon. 
Additionally, DY production with beams of pions and kaons yielded
the structure functions of these unstable particles for the first time.
Also notable in the history of the DY process were the discoveries of the 
$W^\pm$ and $Z^0$ particles in 1983, produced by a      
generalized (vector boson exchange) quark-antiquark annihilation mechanism.

In the mid 1980s some nagging theoretical issues were resolved, with the 
result that by the 1990s the DY process had joined DIS as a
source of quantitative information on the quark structure of hadrons. 
Because the DY process at leading order explicitly involves one antiquark, it 
provides a sensitive measurement of antiquark structure even in kinematic 
regions where quarks dominate in number.

The peaks in Figure~\ref{fige866a} are interlopers in an otherwise well 
understood spectrum. They are seen along with the DY pairs because the 
$^3S_1$ states of heavy quarkonium have substantial 
decay branches to dileptons. However, the production of heavy quarkonium 
is a strong interaction process. At beam energies well above threshold, gluon 
fusion is the dominant production mechanism.
Unlike the well-characterized DY process, the theoretical description of 
quarkonium production at the low transverse momenta ($p_t$) 
characteristic of fixed-target experiments requires a number 
of model-dependent assumptions~\cite{bfy}.

Although the production mechanisms of quarkonia and the DY continuum differ 
greatly, these two types of processes are often reviewed 
together~\cite{kenyon,lyons,shochet,freudenreich} because the data 
come from the same experiments. We continue that tradition here.
The main emphasis of this review is the application of dilepton 
production as a tool to investigate particular physics topics. 
New experimental work has been carried out in recent 
years by few but prolific collaborations working in the fixed-target 
programs at the CERN SPS accelerator and at Fermilab. This review 
concentrates on 
new experimental results published after the comprehensive review of 
Freudenreich~\cite{freudenreich}.

Lepton-pair production forms a major thrust of the collider 
program at Fermilab. We review those results only as they pertain 
directly to the physics addressed by the fixed-target experiments. 

Section 2 presents an overview of the DY process, highlighting 
significant developments in theory and experiment. Sections 3.1 through 
3.4 present recent and future applications of the DY process to the study 
of the quark structure of hadrons. In the remainder of Section 3, we 
discuss several physics topics related to the nuclear dependence of 
the DY process. Section 4 reviews the status of quarkonium and open 
heavy-flavor production in nuclei 
and their relation to quarkonium suppression as a signal for the formation 
of the quark-gluon plasma (QGP) in high-energy collisions of heavy ions. 
Finally, in Section 5, we look at some of the near-future opportunities 
in dilepton physics.

\noindent{2.  THE DRELL-YAN PROCESS} 

\noindent{2.1 Drell-Yan Process in the Parton Model}

\begin{figure}   
\centerline{  
\psfig{figure=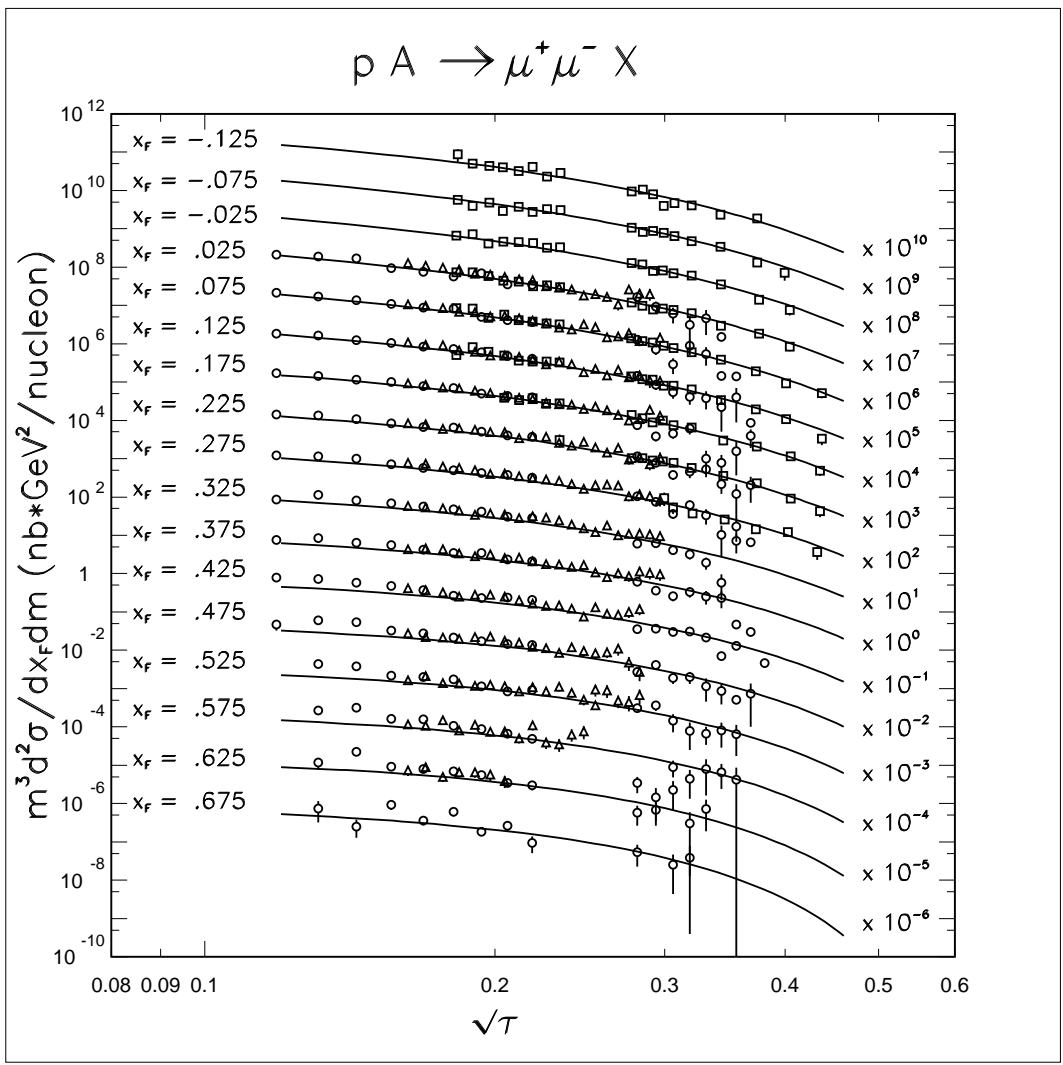,height=5.52in}}
\caption{Proton-induced Drell-Yan production from experiments 
NA3~\cite{na3dy} (triangles) at 
400 GeV/c, E605~\cite{e605} (squares) at 800 GeV/c, and 
E772 (\cite{e772dimu}; PL McGaughey et al, unpublished data) (circles) at 
800 GeV/c. The lines are absolute (no 
arbitrary normalization factor)
next-to-leading order calculations for 
$p + d$ collisions at 800 GeV/c using the                  
CTEQ4M structure functions~\cite{cteq}.}
\label{figsum}                                                     
\end{figure}

The parton model expression for the DY process conveys the essential 
simplicity of the reaction.
\begin{eqnarray}
{d^2\sigma\over dM^2dx_F}={4\pi\alpha^2\over 9M^2s}{1\over (x_1+x_2)}\sum_a
e_a^2[q_a(x_1)
\bar q_a(x_2)+\bar q_a(x_1)q_a(x_2)]. \label{eq:dy}
\end{eqnarray}
Here $q_a(x)$ are the quark or antiquark structure functions of 
the two colliding hadrons evaluated at momentum fractions $x_1$ and 
$x_2$. The sum is over quark flavors.  
In addition, one has the kinematic relations,
\begin{eqnarray}
& &\tau\equiv x_1x_2 = M^2/s,\nonumber \\
& &x_F = x_1-x_2, \label{eq:dykin}
\end{eqnarray}
where $M$ is the invariant mass of the lepton pair and $s$ is the square 
of the center-of-mass energy. The cross section is 
proportional to $\alpha^2$, indicating its electromagnetic character. 
The parton-model scaling properties of the DY process are illustrated in 
Figure~\ref{figsum}, which shows data from experiments at 400 
GeV/c~\cite{na3dy} and at 800 GeV/c (\cite{e605,e772dimu}; PL McGaughey
et al, unpublished data).

In the parton model, the angular distribution of dileptons is
characteristic of the decay of a transversely polarized virtual photon, 
\begin{eqnarray}
{d\sigma \over d\Omega} = \sigma_0 ( 1 + \lambda cos^2 \theta),
\label{eq:dytheta}
\end{eqnarray}
where $\theta$ is the polar angle of the lepton in the virtual photon
rest frame and $\lambda = 1$. Early experimental 
data from both 
pion and proton beams~\cite{kenyon} were consistent with this form
but had large statistical errors.

\begin{figure}
\centerline{
\psfig{figure=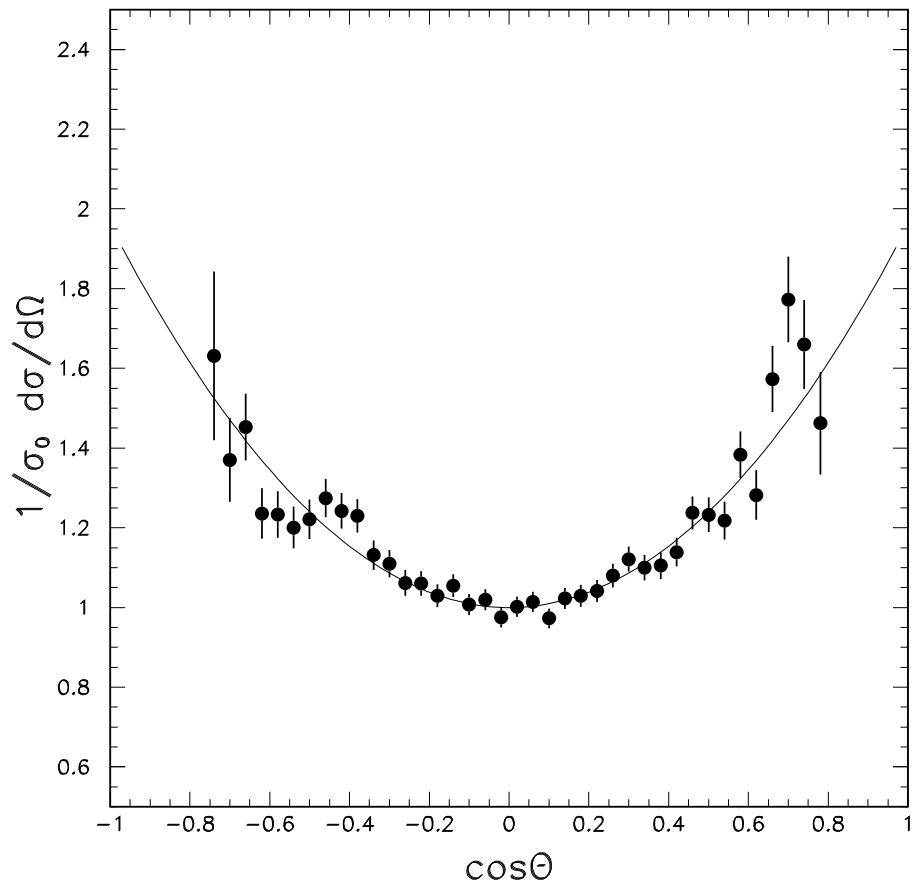,height=5.22in}}
\caption{Drell-Yan angular distribution from Fermilab E772~\cite{QM96}:
$p + Cu$ collisions at 800 GeV/c. The dimuons cover the mass region 
$11\leq M_{\mu^+\mu^-} \leq 17$ GeV/c$^2$ with 
$-0.3\leq x_F\leq 0.8$ and $p_t \leq$ 6 GeV/c. Mean
values for $p_t$, $x_F$, and $M$ are 1.4 GeV/c, 0.16, and
11.9 GeV/c$^2$, respectively. The solid curve is a fit to the data
with the form $1 + \lambda cos^2\theta$, where $\lambda$ is 
$0.96 \pm .04 \pm .06$.}
\label{dyang}
\end {figure}

Recently, E772 has performed a high-statistics study of the angular 
distribution for DY events~\cite{QM96} 
with masses above the $\Upsilon$ family
of resonances. About 50,000 events were recorded from 
800 GeV/c $p + Cu$ collisions, using a copper beam dump
as the target. Figure~\ref{dyang} shows the 
acceptance-corrected angular distribution, 
integrated over the kinematic variables.
Analyzed in the Collins-Soper reference frame~\cite{collins}, the data yield 
$\lambda = 0.96\pm 0.04\pm 0.06$(systematic).

\noindent{2.2 QCD and the Drell-Yan Process} 

Logarithmic scaling violation of the DY cross section 
is expected in leading order 
as for DIS. In the DY process, however, the virtual photon is 
timelike, $M^2=Q^2\ge 0$. Direct experimental confirmation has been 
difficult for the DY process (see the discussion in Freudenreich's 
review~\cite{freudenreich}) for two reasons. First, there is 
the complication 
that the cross section involves the convolution of two structure functions. 
Second, and more importantly, the DY process is experimentally well established 
as the major contribution to the dilepton spectrum only 
for $M\geq 4$ GeV/c$^2$.
This is already a large value of $Q^2$ by DIS standards. Increasing $M$ 
by a factor of five or so to provide enough lever arm to see the logarithmic 
change in $q(x,M^2)$ involves measurement of an exceedingly small 
cross section. Now that DY data has become an integral feature of 
global parton-structure fitting programs, there is no longer much
concern about testing its logarithmic scaling features. 
An excellent summary of the use of DY data in structure function phenomenology 
appears in a recent review~\cite{durham}.

\begin{figure}                       
\centerline{                 
\psfig{figure=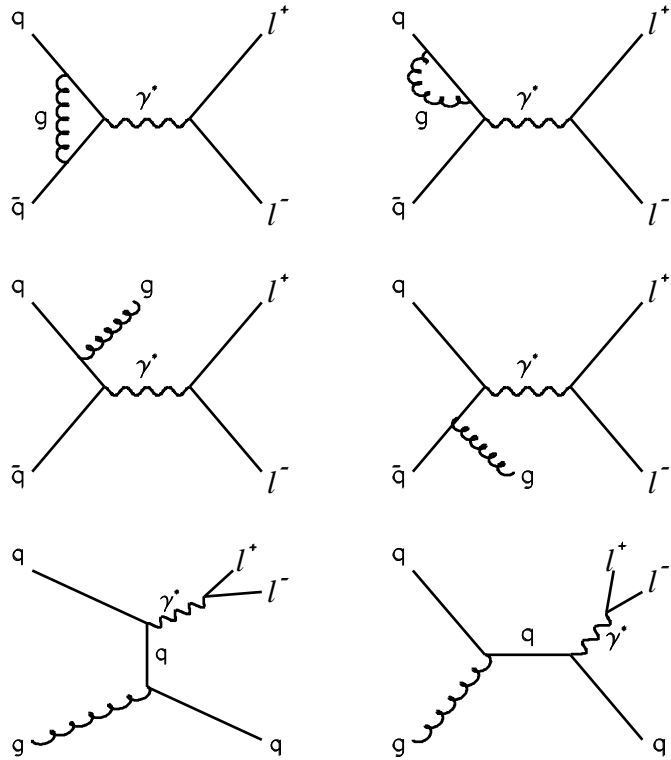,height=5.7in}}
\caption{ Feynman graphs for order $\alpha_s$ corrections to
the Drell-Yan process}
\label{dynlo}                                
\end{figure}

The importance of next-to-leading order (NLO; terms 
proportional to $\alpha_s$) QCD corrections to the DY process has been 
well known since the mid-1980s~\cite{shochet,altarelli}. Historically, two 
experimental features demanded theoretical improvement: first,
the experimental cross section was about a factor of 
two larger than the parton-model value, and second, 
the distribution of dilepton transverse momenta extended to much larger 
values than are characteristic of the convolution of intrinsic parton 
momenta. Corrections to the DY mechanism at NLO
largely fix these two problems. As shown 
in Figure~\ref{dynlo}, this comes at the 
expense of involving diagrams that seem to eliminate the simple parton 
interpretation of $q\bar q$ annihilation. Nevertheless, for most 
interpretations of the DY process, it is reasonably accurate to think 
in terms of parton-level $q\bar q$ annihilation with a modest 
increase in cross section.

QCD provides a procedure for calculating higher-order corrections to the 
DY process as well~\cite{rijken}. In practice,
NLO corrections are normally employed in 
QCD phenomenology. Figure~\ref{figsum} shows the typical agreement found 
when NLO calculations are compared $-$ without normalization $-$ to data.

Including higher-order QCD corrections to the DY process~\cite{oakes,collinsa}
results in the more complicated form of 
the angular distribution,
\begin{eqnarray}
{d\sigma \over d\Omega} \propto 1 + \lambda cos^2 \theta + 
\mu sin 2\theta cos \phi + {\nu \over 2} sin^2 \theta cos 2 \phi, 
\label{eq:dyangdist1}
\end{eqnarray}
where $\phi$ is the azimuthal angle and $\lambda$,
$\mu$, and $\nu$ are angle-independent parameters. NLO
calculations predict~\cite{chiappetta}
small deviations from $1 + cos^2 \theta$ ($\leq 5\%$) for \pt
below 3 GeV/c. The relevant scaling parameter for the magnitude
of these deviations is $p_t / Q$, implying that NLO 
corrections become important when $p_t \simeq Q$. 
A relation, $1 -\lambda -2\nu = 0$, developed by Lam \& Tung~\cite{lam},
is analogous to the Callan-Gross relation in DIS. Measurements with
pion beams at CERN~\cite{NA10angdist} and at Fermilab~\cite{E615a}
have shown that the Lam-Tung relation is clearly violated at large $p_t$.

Pion-induced DY experiments have unexpectedly shown that
transverse photon polarization changes to
longitudinal ($\lambda \simeq -1$) at 
large $x_F$~\cite{NA10angdist,E615a,NA3angdist,E615}.
The \xf dependence of $\lambda$ is qualitatively consistent with 
a higher-twist model originally proposed by Berger \& 
Brodsky~\cite{berger,bergera}. However, the quantitative 
agreement is poor. The model's
basis can be described as follows.
As \xf of the muon pair approaches unity, the Bjorken-$x$ (momentum
fraction) of the annihilating projectile parton must also be
near unity. Thus, the whole pion 
contributes to the DY process. This can be treated with perturbation 
theory, with the result that the transverse polarization of the
virtual photon becomes longitudinal. The angular distribution at
large \xf becomes

\begin{eqnarray}
{d\sigma \over d\Omega} \propto (1-x)^2 (1 + \lambda cos^2 \theta) + 
\alpha sin^2 \theta, \label{eq:dyangdist12}
\end{eqnarray}

\noindent where $\alpha$ is $\propto$ $p_t^2 / Q^2$.

Eskola et al~\cite{eskola} have shown that an improved treatment
of the effects of nonasymptotic kinematics greatly improves
quantitative agreement with the $\lambda$ values from the pion data. 
Brandenburg et al~\cite{brandenburg} have extended the higher
twist model to specifically include pion bound-state effects.
They predict values for $\lambda$, $\mu$ and $\nu$ that are in
good agreement with the pion data at large $x_F$. Unfortunately,
the results are quite sensitive to the choice of the pion Fock
state wave functions, which are not well constrained by experimental
data.

At this time, no comprehensive theory covering the full kinematic 
range of the experiments is available, nor is the observed violation 
of the Lam-Tung rule understood. 

\noindent{2.3 Factorization and Nuclear Size Effects}

General discussions of factorization in the DY process and other QCD 
processes may be found in reviews by Collins \& Soper~\cite{collinssoper2} 
and Sterman et al~\cite{sterman}.

A crucial requirement for the application of the DY process for the study of 
the quark structure of nuclei and nucleons (where most targets are actually 
nuclei) is that the factorized form of the cross 
section expressed by Equation~\ref{eq:dy} (or its QCD-corrected equivalent) 
remains valid for nuclei. Specifically, one wants assurance that
$q_a(x,M^2)$ is the same for a proton that has traversed the diameter of 
an atomic nucleus as it is for a free proton. Fortunately, this 
problem was analyzed extensively in the 1980s, particularly for the DY 
process~\cite{bbl,css,bodwin}. Within limits given approximately by 
\begin{eqnarray}
Q^2\geq  x_2m_NL_A\langle l_{\perp}^2\rangle, \label{eq:targetl}
\end{eqnarray}
for hadron-nucleus collisions, initial-state interaction effects can be 
ignored for observables integrated over dilepton transverse momenta. 
In Equation~\ref{eq:targetl}, $x_2$ is 
the parton momentum fraction for a nucleus 
of mass $A$, $m_N$ is the nucleon mass, and $l_{\perp}\approx 0.3$ GeV/c is 
the average momentum exchange in a quark-nucleon collision.
With a path length $L_A = 2R_0A^{1/3}$, Equation~\ref{eq:targetl} 
becomes $Q^2\geq x_2A^{1/3}$ 
GeV$^2$/c$^2$, a condition readily met for all nuclei.
New nuclear dependence measurements, discussed in Sections 3.5-3.7,
provide quantitative evidence of the absence of substantial initial-state 
interaction effects.

\noindent{2.4 Experimental Detection of Continuum Lepton Pairs}

The DY process was anticipated theoretically in 1970~\cite{drell}. 
However quantitative experiments had to wait until the late 1970s for 
the development of techniques permitting the measurement of 
picobarn cross sections in the presence of background processes many 
orders of magnitude larger. The most successful high-luminosity DY 
experiments have exploited inclusive dimuon detection in magnetic 
spectrometers whose apertures are filled with hadron absorbers.
Hadron absorption near the target also 
reduces the backgrounds of muons from pion and kaon decays.
Freudenreich's 1990 review~\cite{freudenreich} of lepton-pair 
production includes detailed descriptions of spectrometers built and 
operated during the 1970s and early 1980s. During the late 1980s and 
1990s, two dimuon spectrometers remained operational at Fermilab and 
two at the CERN SPS. This review concentrates on data from these 
instruments. (The HELIOS collaboration at the SPS focused 
on the mass region below the $J/\psi$; hence, their work is not covered 
by this review.)

\begin{figure}                
\centerline{
\mbox{\rotate[r]{\epsfysize=5.5in\epsffile{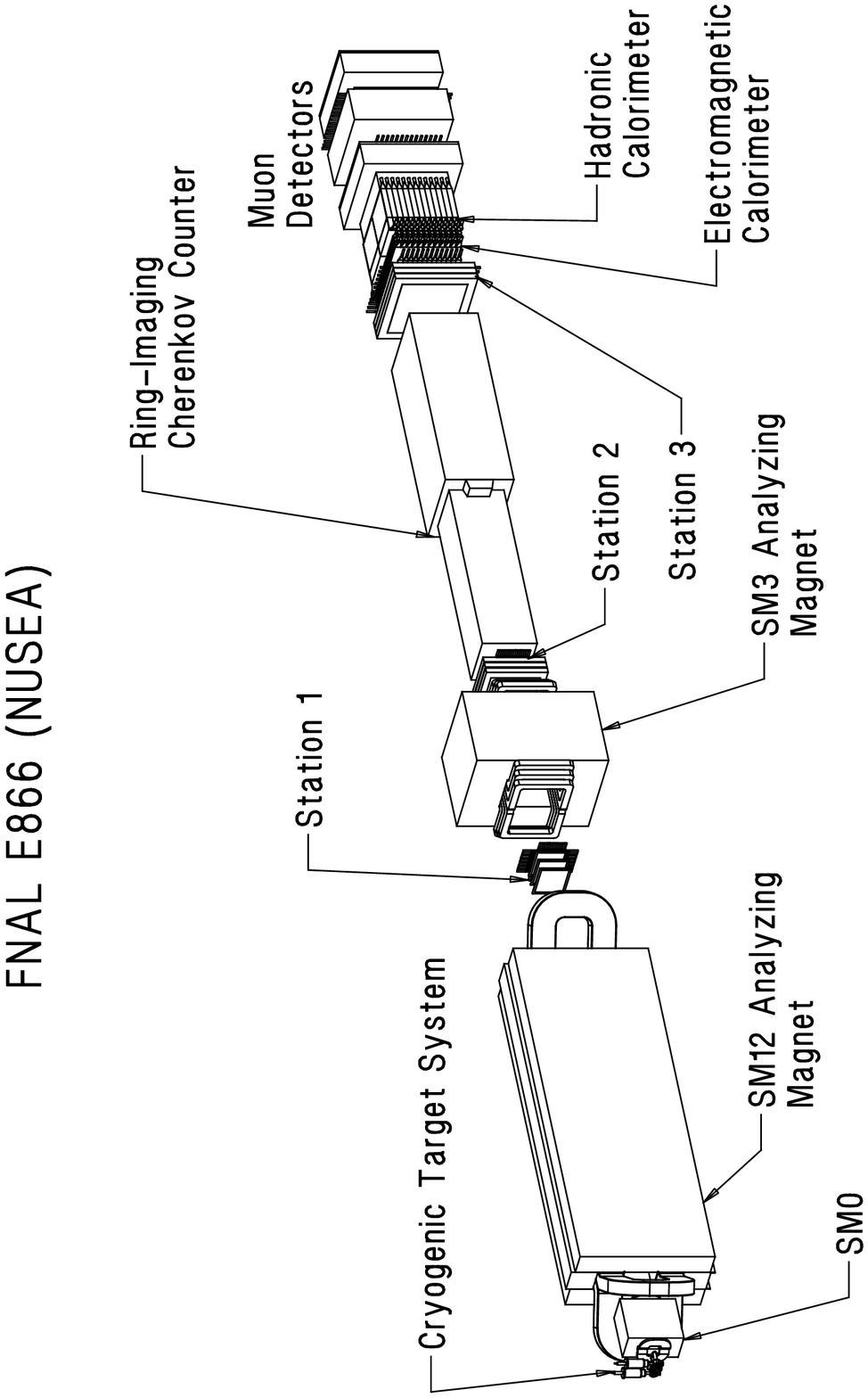}}}}
\caption{ Schematic layout of the Meson-East focusing
spectrometer at Fermilab.}
\label{fige866b}         
\end{figure}

At Fermilab, a series of three experiments $-$ E772, 
E789, and E866 $-$ used differing configurations of the large spectrometer 
located on a direct 800-GeV/c proton beam line in the Meson-East 
experimental area. The spectrometer (Figure~\ref{fige866b}) was originally 
constructed for E605, which studied dimuon production 
for $M_{\mu^+\mu^-}\geq$ 7 GeV/c$^2$. 
The first two dipole magnets focus high-mass pairs into the spectrometer 
acceptance, thus avoiding the beam dump contained in the second magnet. 
The dump absorbs not only the beam but also the 
enormous flux of low-\pt secondaries from the target. The third large analyzing 
magnet provides a \pt kick of about 1 GeV/c for momentum analysis by the tracking 
system. The spectrometer is capable of operating at luminosities in the range 
$10^{35}cm^{-2}sec^{-1}$ with 10\% interaction length targets and detection 
acceptances of $\sim 1-2\%$ for the DY process. E772, E789, and E866 also exploited 
the copper beam dump as a second target~\cite{QM96,kowitt}. The beam 
dump ``target'' 
has the advantage of a more open acceptance, permitting the measurement of angular 
distributions of muon pairs over wide range of decay angles.

The E705 collaboration at Fermilab operated an open-aperture (no hadron 
absorber) spectrometer which contributed important data to the study of 
the \J and \JP resonances and their decays~\cite{antoniazzi}.

The muon-pair spectrometer operated most recently at CERN was built 
originally for experiment NA10~\cite{na10}. It consists 
of a large air-core toroid 
magnet with tracking chambers following the hadron absorber. In recent 
years, with substantial changes in its configuration, it has made unique 
contributions to the study of \J production in relativistic heavy-ion 
collisions at the SPS~\cite{na50}, as well as to the 
determination of the $\bar d/\bar u$ asymmetry of the proton~\cite{na51}. 

\noindent{2.5 Backgrounds in Inclusive Dimuon Experiments}

The E866 spectrum (Figure~\ref{fige866a})
illustrates the typical experimentally ``safe'' regions for DY muon pairs 
in fixed-target experiments with beams in the few hundred GeV/c range. Below 
the \J, a number of potential backgrounds, including the semileptonic 
decay of charmed hadrons, make the interpretation of the continuum more 
complex. Additionally, depending on details of the target and 
absorber configuration, pion and kaon decays can affect the low-mass 
spectrum. The data can be corrected for the latter contribution via 
subtraction of properly normalized like-sign muon pairs. But charm pair 
decay, where this technique does not work, presents a more serious problem 
and ultimately limits the kinematic region that can be exploited.
Ideally, the charm-pair decay contribution could be separated by use of a 
microvertex detector. To date, however, such a detector has not been 
combined with closed-aperture dimuon detection. The most systematic 
study of the contribution of charm production to the dimuon spectrum has 
been performed recently by the NA50 collaboration~\cite{na50} in 
connection with measurements of \J 
suppression in heavy-ion collisions. In NA50, the charm-pair decay continuum 
is evaluated with the help of simulations of all contributing processes 
in the mass region just below the $J/\psi$.

\noindent{3. DRELL-YAN PROCESS AND PARTON DISTRIBUTIONS}

\noindent{3.1 $\bar d/\bar u(x)$ Asymmetry of the Proton}

From neutrino-induced DIS experiments, it is known that the strange-quark
sea in the nucleon is roughly a factor of two less than the up or down
quark sea~\cite{cdhs}. The lack of SU(3) flavor 
symmetry in the nucleon sea has been 
attributed to the much heavier mass of the strange quark. Until recently,
it had been assumed that the distributions of $\bar u$ and $\bar d$
quarks were identical. Although the equality of $\bar u$ and $\bar d$
in the proton is not required by any known symmetry, it is a plausible
assumption for sea quarks generated by gluon splitting. Because 
the masses of the 
up and down quarks are small compared to the confinement scale, nearly equal 
numbers of up and down sea quarks should result.

The assumption of $\bar u(x) = \bar d(x)$ can be tested by measurements
of the Gottfried integral~\cite{gott}, defined as
\begin{equation}
I_G = \int_0^1 \left[F^p_2 (x,Q^2) - F^n_2 (x,Q^2)\right]/x~ dx =
{1\over 3}+{2\over 3}\int_0^1 \left[\bar u_p(x)-\bar d_p(x)\right]dx,
\label{eq:3.1}
\end{equation}
where $F^p_2$ and $F^n_2$ are the proton and neutron structure
functions measured in DIS experiments. The second step in 
Equation~\ref{eq:3.1} follows from the assumption of nucleon 
charge symmetry (Section 3.3).
Under the assumption of a symmetric sea, $\bar u$ = $\bar d$,
the Gottfried Sum Rule (GSR)~\cite{gott}, $I_G
= 1/3$, is obtained. Several early attempts~\cite{slacgsr,emcgsr,bcdmsgsr}
to test the GSR were inconclusive
because the measurements did not reach small enough $x$, where large 
contributions to the integral are expected. Nevertheless, 
Field \& Feynman~\cite{ff} interpreted the early SLAC 
data~\cite{slacgsr}, which gave $I_G(0.02 \to 0.8) = 0.20 \pm 0.04$,
as an indication that the GSR is violated with $\bar d(x) > \bar u(x)$.
They suggested that the Pauli-blocking effect suppressed $g \to u \bar u$ 
relative to $g \to d \bar d$, since protons contain two $u$-valence quarks
and only one $d$-valence quark.

The most accurate test of the GSR was reported in 1991 by the New Muon 
Collaboration (NMC)~\cite{nmc1}, which measured $F^p_2$ and $F^n_2$ over the 
region $0.004 \le x \le 0.8$. They determined the Gottfried integral to be 
$ 0.235\pm 0.026$, significantly below 1/3. This surprising result has
generated much interest.

Although the violation of the GSR can be explained by
assuming unusual behavior of the parton distributions at very
small $x$~\cite{martin},
a more natural explanation is to abandon the assumption
$\bar u = \bar d$.  Specifically, the NMC result implies
\begin{equation}
\int_0^1 \left[\bar d(x) - \bar u(x)\right] dx = 0.148 \pm 0.039.
\label{eq:3.2}
\end{equation}
We emphasize that only the integral of $\bar d -\bar u$ was deduced
from the DIS measurements. The $x$ dependence of $\bar d - \bar u$ remained
unspecified.

The proton-induced DY process provides an
independent means to probe the flavor asymmetry of the nucleon sea~\cite{es}.
An important advantage of the DY process is that the $x$ dependence of 
$\bar d / \bar u$ can be determined.
It is interesting to note that, as early as 1981, Fermilab E288~\cite{ito} 
reported evidence for a $\bar d/\bar u$ asymmetry, 
based on a measurement of the $p + d$ DY cross
section. However, this interpretation depended sensitively on assumptions
about the shape of the valence quark distributions and was 
not conclusive. Later, the 
Fermilab E772 collaboration~\cite{plm}
compared the DY yields from isoscalar targets with that from a neutron-rich 
(tungsten) target, and constraints on the nonequality of $\bar u$ and 
$\bar d$ in the range $0.04 \leq x \leq 0.27$ were set. 
More recently, the CERN experiment NA51~\cite{na51}
carried out a comparison of the DY muon pair yield from hydrogen and
deuterium using a 450 GeV/c proton beam. They found that
$\bar u / \bar d = 0.51 \pm 0.04 \pm 0.05$ at $\langle x \rangle =0.18$, a 
surprisingly large difference between the $\bar u$ and $\bar d$.

\begin{figure}
\center
\hspace*{0.4in}
\psfig{figure=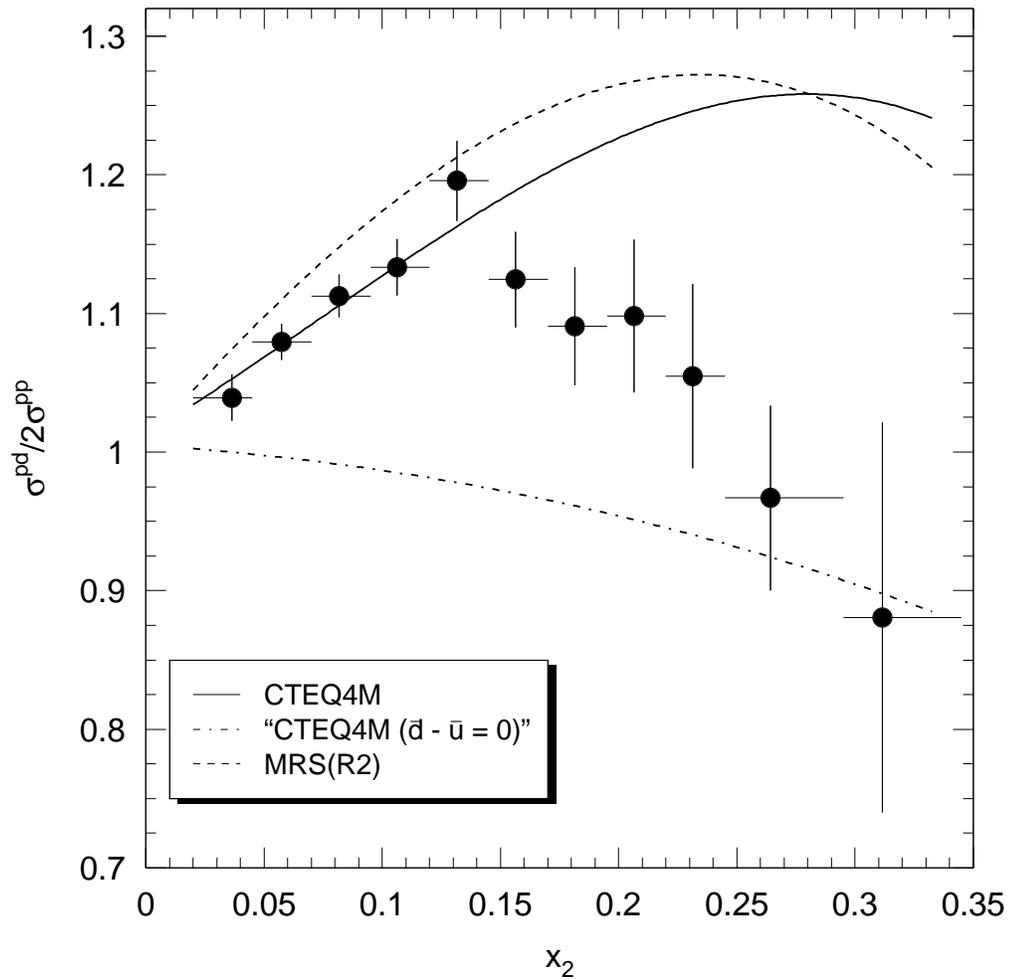,height=5.22in}
\caption{The ratio $\sigma^{pd}/2\sigma^{pp}$ of Drell-Yan cross
sections {\em vs.} $x_{2}$ from Fermilab E866~\cite{e866}. 
The curves are next-to-leading order
calculations, weighted by acceptance, of the Drell-Yan cross section 
ratio using the CTEQ4M~\cite{cteq} and MRS(R2)~\cite{mrs} parton distributions. 
In the lower
CTEQ4M curve, $\bar{d} - \bar{u}$ was set to 0 to simulate a symmetric sea.}
\label{fig:3.1}
\end{figure}

A DY experiment (E866), aiming at higher statistical accuracy and wider 
kinematic coverage than NA51, was recently completed~\cite{e866}
at Fermilab. This experiment also measured the 
DY muon pairs from 800-GeV/c protons interacting with liquid deuterium and hydrogen targets.
The acceptance of the spectrometer
was largest for $x_F = x_1 - x_2 > 0$. In this kinematic regime the DY
cross section is dominated by the annihilation of a beam quark with a target
antiquark. The DY cross section ratio
at large $x_F$ is approximately given as 
\begin{equation}
{\sigma_{DY}(p+d)\over 2\sigma_{DY}(p+p)} \approx
{1\over 2} \left(1+{\bar d(x_2)\over \bar u(x_2)}\right).
\label{eq:3.3}
\end{equation}
The ratio is unity when  $\bar d = \bar u$.
Figure~\ref{fig:3.1} shows that the E866 measurement of this ratio 
clearly exceeds unity for an appreciable range in $x_2$.

\begin{figure}
\center
\hspace*{0.5in}
\psfig{figure=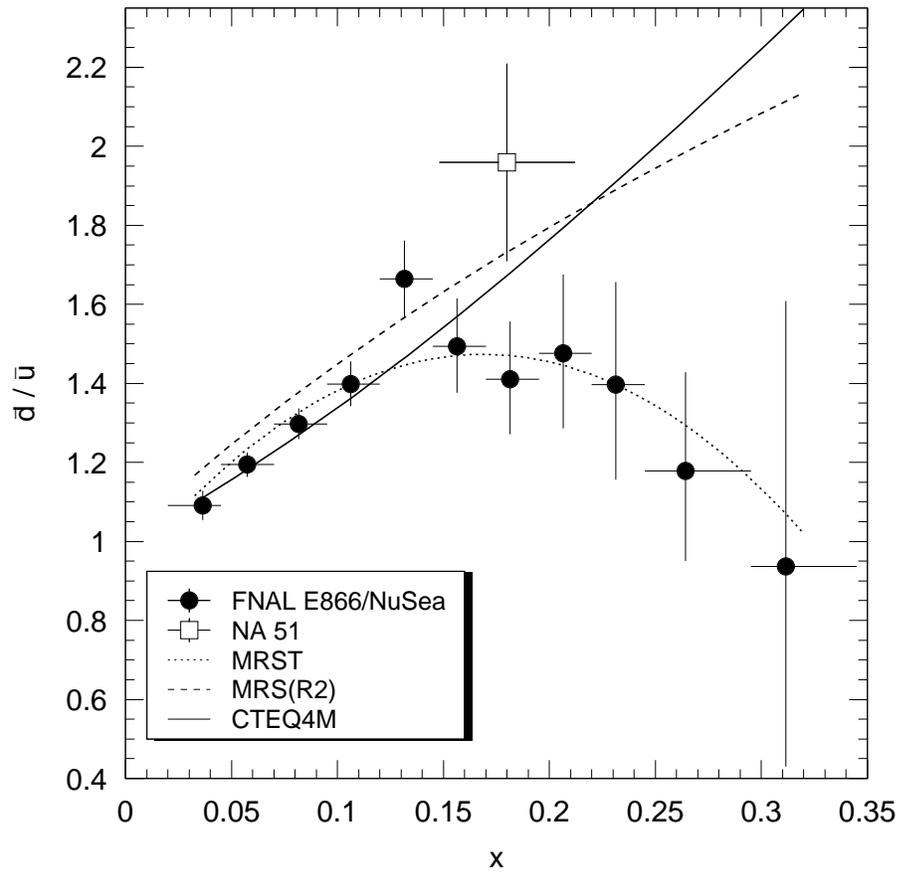,height=4.68in}
  \caption{The ratio of $\bar{d}/\bar{u}$ in the proton as a function
  of $x$ extracted from the Fermilab E866~\cite{e866} cross section ratio. The
  curves are from various parton distributions.  
  Also shown is the result
  from NA51~\cite{na51}, plotted as an open box.}
\label{fig:3.2}
\end{figure}

Using an iterative procedure~\cite{e866}, values for 
$\bar d/ \bar u$ were extracted by the E866 collaboration at
$Q^2 = 54$ GeV$^2$/c$^2$. These are shown in Figure~\ref{fig:3.2} along 
with the NA51 measurement.
For $x < 0.15$, $\bar d/\bar u$ increases linearly with $x$ and is in
good agreement with the CTEQ4M~\cite{cteq} and MRS(R2)~\cite{mrs} parameterizations. 
However, a distinct feature of the data, not seen in either 
parameterization, is the
rapid decrease toward unity of $\bar{d}/\bar{u}$ beyond
$x=0.2$\@. 

\begin{figure}
\center
\hspace*{0.5in}
\psfig{figure=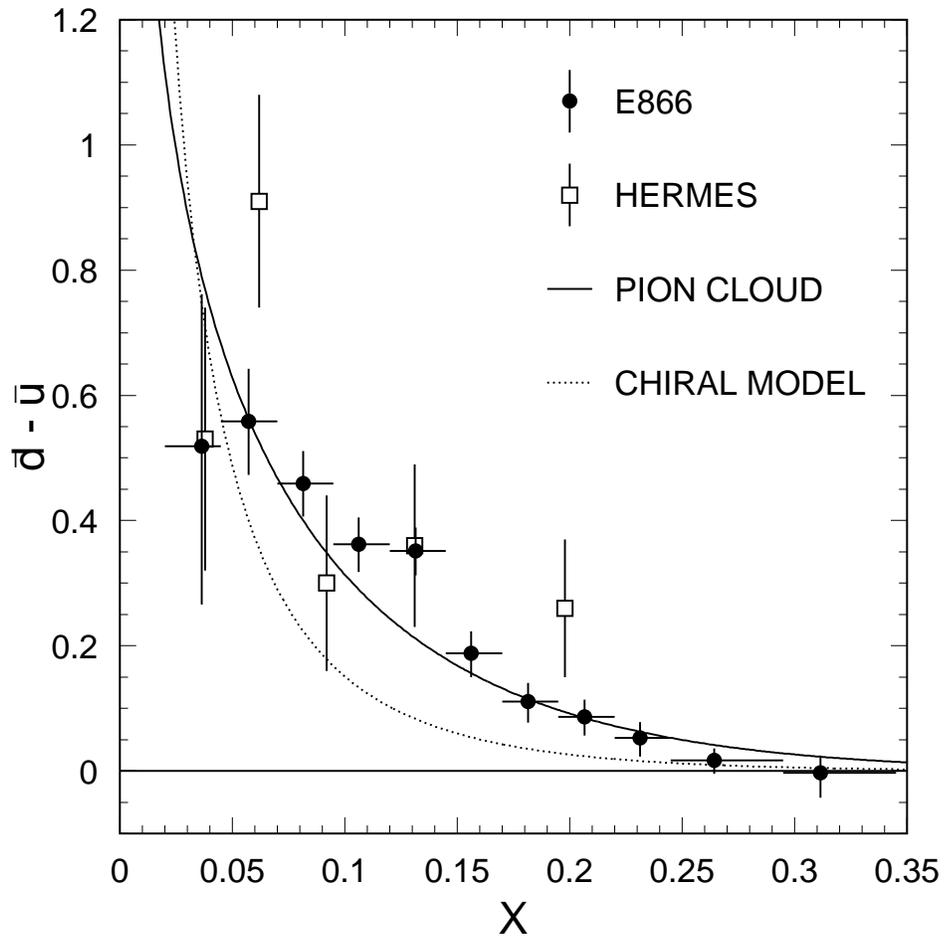,height=5.4in}
\caption{Comparison of the E866~\cite{e866} $\bar d - \bar u$ results at $Q^2$ =
54 GeV$^2$/c$^2$ with the predictions of pion-cloud and chiral models 
as described in the text. The data from HERMES~\cite{hermes} are also shown.}
\label{fig:3.3}
\end{figure}

The $\bar d / \bar u$ ratio, along with the
CTEQ4M values for $\bar d + \bar u$, was used to obtain
$\bar d - \bar u$ over the region $0.02 < x < 0.345$
(Figure~\ref{fig:3.3}). Being a flavor nonsinglet quantity,
$\bar d(x) - \bar u(x)$ is decoupled from gluon splitting.
From the results shown in Figure~\ref{fig:3.3}, one can
obtain an independent determination~\cite{peng} of the integral of 
Equation~\ref{eq:3.2}.
E866 finds $0.100 \pm 0.007
\pm 0.017$, consistent with, but roughly $2/3$ of the value deduced by NMC. 

Very recently, the HERMES collaboration reported a semi-inclusive
DIS measurement of charged pions from hydrogen and deuterium 
targets~\cite{hermes}.
Based on the differences between charged-pion yields from the two targets,
the ratio $(\bar d - \bar u)/ (u -d)$ is determined
in the kinematic range, $0.02 < x < 0.3$ and 
1 GeV$^2$/c$^2 < Q^2 <$ 10 GeV$^2$/c$^2$. The HERMES results for
$\bar d - \bar u$, shown in Figure~\ref{fig:3.3}, are consistent with
the E866 results obtained at significantly higher $Q^2$.

\begin{figure}
\center
\hspace*{0.5in}
\psfig{figure=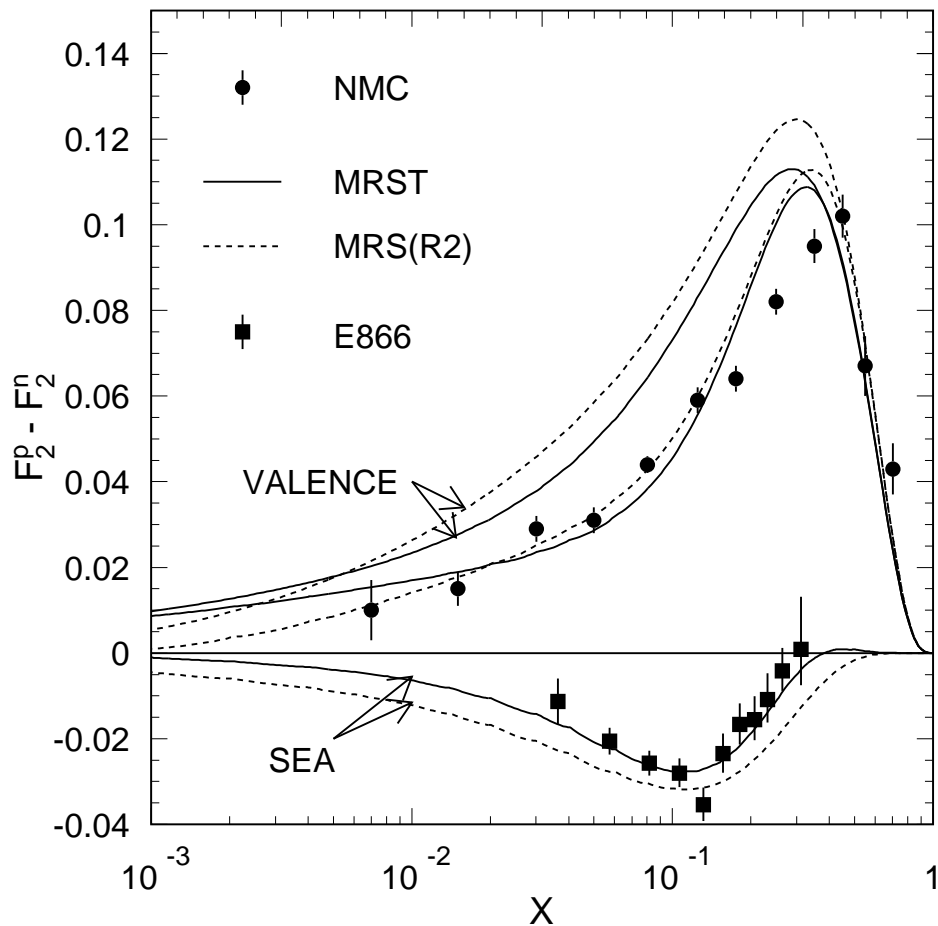,height=5.4in}
\caption{$F^p_2 - F^n_2$ as measured by NMC~\cite{nmc1}
at $Q^2$ = 4 GeV$^2$/c$^2$ compared with
predictions based on the MRS(R2)~\cite{mrs} and MRST~\cite{mrst} 
parameterizations. Also shown are the E866 
results~\cite{e866}, evolved to $Q^2$ = 4 GeV$^2$/c$^2$, for 
the sea-quark contribution to $F^p_2
- F^n_2$. For each prediction, the top (bottom) curve is the valence
(sea) contribution and the middle curve is the sum of the two.}
\label{fig:3.4}
\end{figure}

The E866 data clearly affect the current parameterization of the
nucleon sea. The most recent 
structure functions of Martin et al~\cite{mrst}(MRST)
included the E866 data in their global fit and 
are very different from the previous
MRS(R2) parameterization (Figure~\ref{fig:3.2}).
What is not so obvious is that the E866 data
also affect the parameterization of the valence-quark distributions.
Figure~\ref{fig:3.4} shows the NMC data for $F_2^p
- F_2^n$ at $Q^2$ = 4 GeV$^2$/c$^2$, along with the fits of MRS(R2) and
MRST. It is instructive to
decompose $F_2^p(x) - F_2^n(x)$ into contributions from 
valence and sea quarks:
\begin{eqnarray}
F_2^p(x) -F_2^n(x) = 
 {1 \over 3} x \left[u_v(x) - d_v(x)\right] + {2 \over
3} x \left[\bar u(x) - \bar d(x)\right].
\end{eqnarray}
As shown in Figure~\ref{fig:3.4}, the E866 data
provide a direct determination of the sea-quark contribution to $F_2^p
- F_2^n$. In order to preserve the fit to $F_2^p - F_2^n$, the MRST
parameterization of $u_v - d_v$
is significantly different from that of MRS(R2). Neither set reproduces 
$F_2^p- F_2^n$ in the range $x=0.2-0.3$.

We now turn to the origin of the $\bar d / \bar u$ 
asymmetry~\cite{peng,kumano0}.  As
early as 1983, Thomas~\cite{thomas} pointed out that the virtual pions
that dress the proton will lead to an enhancement of $\bar d$ relative
to $\bar u$ via the (nonperturbative) 
``Sullivan process.''
Sullivan~\cite{sullivan} previously showed that in DIS 
virtual mesons scale in the Bjorken limit and contribute to the
nucleon structure function.  Following the publication of the NMC
result, many 
papers~\cite{henley,kumano,signal,hwang,szczurek,koepf,ehq,cheng1,szczurek2}
treated virtual mesons as the origin of the $\bar d/\bar u$ asymmetry
(see~\cite{kumano0} for a recent review).
Here the $\pi^+(\bar d u)$ cloud, dominant in the 
process $p\rightarrow\pi^+ n$, leads to an excess of $\bar d$ sea.

A different approach for including the effects of virtual mesons has
been presented by Eichten et al~\cite{ehq} and
further investigated by other authors~\cite{cheng1,szczurek2}. In 
chiral perturbation theory, the relevant degrees of
freedom are constituent quarks, gluons, and Goldstone bosons. In
this model, a portion of the sea comes from the couplings of Goldstone
bosons to the constituent quarks, such as $u \to d \pi^+$ and $d \to u
\pi^-$. The excess of $\bar d$ over $\bar u$ is then simply due to the
additional valence $u$ quark in the proton. 

The $x$ dependences of $\bar d - \bar u$ and $\bar d / \bar u$ obtained
by E866 provide important constraints for theoretical models.
Figure~\ref{fig:3.3} compares $\bar d(x) - \bar u(x)$ from E866
with a virtual-pion model calculation, following the procedure detailed
by Kumano~\cite{kumano}. A dipole form, with $\Lambda = 1.0$ GeV for 
the $\pi N N$ form factor and $\Lambda = 0.8$ GeV for 
the $\pi N \Delta$ form factor,
was used. $\Lambda$ is the cutoff parameter for the pion form factor.
Figure~\ref{fig:3.3} ({\it dotted curve}) also shows the predicted
$\bar d - \bar u$ from the chiral model.
We follow the
formulation of Szczurek et al~\cite{szczurek2} to calculate
$\bar d(x) - \bar u(x)$ at $Q^2$ = 0.25 GeV$^2$/c$^2$ and then 
evolve the results
to $Q^2$ = 54 GeV$^2$/c$^2$. The chiral model places 
more strength at low $x$
than does the virtual-pion
model. This difference reflects the fact
that the pions are softer in the chiral model, since they are coupled
to constituent quarks, that carry only a fraction of the
nucleon momentum. The $x$ dependence of the E866 data favors the
virtual-pion model over the chiral model, suggesting that
correlations between the chiral constituents should be taken into
account.

The chiral and the meson-cloud models both predict that the $\bar u$ and
$\bar d$ quarks will carry negligible amount of the proton's 
spin~\cite{peng,ehq,cheng2}, a
prediction that remains to be tested.
Another interesting consequence of the meson-cloud model
is that the $s$ and $\bar s$ distributions
in the proton could have very different shapes, even though the net amount
of strangeness in the proton vanishes~\cite{ji,brodsky}. By 
comparing the $\nu$ and $\bar \nu$
induced charm production, the CCFR 
collaboration found that 
the $s$ and $\bar s$ distributions are very similar~\cite{bazarko}. 
Dimuon production experiments using $K^\pm$ beams might provide an independent
determination of the $s$/$\bar s$ ratio of the proton. 

{\noindent 3.2 Polarized Drell-Yan and $W^\pm$ Production}

The spin structure of the nucleon has been extensively investigated by 
polarized DIS~\cite{anselmino,hughes} during the past 10 years. Much of 
the excitement in this field has come from the finding that quarks
contribute a surprisingly small fraction to the nucleon's
total helicity. As a result, considerable interest has been centered 
on QCD processes in hadronic collisions, where polarized gluon and antiquark 
effects are directly manifested. 

Polarized DY 
production and $W^{\pm}$ production have great potential for providing 
qualitatively new information about antiquark 
polarization.\footnote{Quarkonium production has also been considered as 
a means of determining gluon polarization. However, unresolved issues 
about the mechanism of quarkonium production (see Section 4) prevent 
its application as even a qualitative measure of $\Delta G$, at least in the  
early stages of polarized hadronic collisions.}
Because polarized proton collisions will become feasible at the RHIC 
facility starting in 2001, it is timely to review the type of 
information that such studies will provide.

The longitudinal spin asymmetry (requiring polarization of both protons)
in the DY process is, in leading order, 
given by~\cite{close},
\begin{eqnarray}
A_{LL}^{DY}(x_1,x_2)={\sum_a e_a^2[\Delta q_a(x_1) \Delta \bar
q_a(x_2)+\Delta\bar q_a(x_1) \Delta q_a(x_2)]\over\sum_a e_a^2[q_a(x_1)
\bar q_a(x_2)+\bar q_a(x_1)q_a(x_2)] }, \label{eq:ALL0} 
\end{eqnarray}
with $\Delta q_a\equiv q_a^+-q_a^-$.  The superscripts refer to parton
spin projections parallel ($+$) or antiparallel ($-$) to the proton's
spin projection.\footnote{Next-to-leading order QCD corrections to 
Equation~\ref{eq:ALL0} have been evaluated by several 
authors~\cite{ratcliffe,gehrmann,kamal}. 
They are generally small in most of the DY kinematic range.} 
One can simplify this equation 
by choosing the kinematic region $x_F\geq 0.2$, where a single term,
$u(x_1)\bar u(x_2)$, dominates 
the denominator of Equation~\ref{eq:ALL0} (See Section 3.1). The 
equation thus becomes
\begin{eqnarray}
A_{LL}^{DY}(x_1,x_2) & \approx & A_{LL}^{pDIS}(x_1)\times{\Delta\bar 
u\over \bar u}(x_2) \label{eq:ALL1},
\end{eqnarray}
where the asymmetry measured in DIS on polarized protons is 
$A_{LL}^{pDIS}(x)=g_1(x)/F_1(x)$.
Thus, the helicity asymmetry in polarized p-p collisions is a measure of 
the antiquark polarization at $x_2$ in one proton, with the other proton 
providing a quark at $x_1$ with known polarization~\cite{close,moss}.
(This derivation also  
requires that the DIS structure function, $g_1^n(x)$, be approximately
zero, a condition  
well met for $x\geq 0.2$.)

A closely related reaction, the production of $W^{\pm}$, has been analyzed 
theoretically~\cite{bourrely}.  The parity-violating 
nature of $W^{\pm}$ production leads to some essential differences. 
First, only one of the two beams needs to be polarized. Second,  
$W^{\pm}$ leptonic decay
produces a charged lepton, which can be detected, and a neutrino, which 
cannot. Thus, the measurement necessarily integrates 
over some range of $x_1$ and $x_2$.
Using the kinematics treated previously for DY
production, one finds, for $W$s produced at negative $x_F$ (opposite to 
the direction of the polarized beam ($x_1$)), 
\begin{eqnarray}
A_L^{W^+}\approx -{\Delta\bar d \over\bar d}(x_2),\ {\rm and}\ \ \
A_L^{W^-}\approx -{\Delta\bar u\over\bar u}(x_2).  \label{eq:ALW1}
\end{eqnarray} 
This is similar to Equation~\ref{eq:ALL1} but without the
quark polarization factor multiplying the antiquark asymmetry.
For $W$s produced at positive $x_F$, one finds, 
\begin{eqnarray}
A_L^{W^+}\approx {\Delta u \over u}(x_2),\ {\rm and}\ \ \
A_L^{W^-}\approx {\Delta d\over d}(x_2).  \label{eq:ALW2} 
\end{eqnarray}

In the DY process (or in $Z^0$ production, but 
not $W^{\pm}$ production), it is also possible to
measure a new structure function,
called transversity, which is a correlation between quark 
momentum and its perpendicular spin component~\cite{ralston}. 
The transversity is not measurable in inclusive DIS~\cite{jaffe}.
It is measurable, in 
principle, in collisions of polarized protons whose spins are aligned 
perpendicular to the plane of dilepton detection. A nonzero transverse 
spin correlation in the DY 
process would clearly require both quark and antiquark transversities to be 
nonzero. A theoretically ideal, but currently impossible, experiment 
would be the measurement of the transverse spin correlation and hence 
the quark transversity of the proton via polarized $p\bar 
p$ collisions.\footnote{An estimate of the transversity asymmetry in the bag 
model~\cite{jaffe} indicates that it peaks at large $x$, as does the 
helicity asymmetry. Polarized $p\bar
p$ collisions at large $x$ would thus yield a large experimental 
asymmetry.}

Studies of both continuum DY and $W^{\pm}$ production in polarized p-p collisions 
are planned at RHIC~\cite{bunce}. The scaling 
properties of the DY cross section, 
\begin{eqnarray} 
d^2\sigma/d\sqrt\tau d x_F\propto 1/s,
\label{eq:scale} 
\end{eqnarray}
combined with the kinematic relation, Equation~\ref{eq:dykin}, greatly favors the
lowest beam energy consistent with the production of DY pairs in the
``safe'' region. Thus, detection of muon pairs with $x_1 = 0.25$, $x_2 = 
0.4$ (M = 5 GeV/c$^2$) at $\sqrt s = 50$ GeV has 16 times higher cross section 
than detection of 20-GeV/c$^2$ pairs at $\sqrt s = 200$ GeV. Because of its 
high threshold, $W^{\pm}$ is feasible only at the highest proton energy 
of RHIC, $\sqrt s = 500$ GeV.

\noindent{3.3 Charge Symmetry Violation in Parton Distributions}

Charge symmetry is believed to be well respected in strong interaction.
Extensive experimental searches for charge symmetry violation (CSV) 
effects in various nuclear processes reveal an amount on
the order of $1\%$~\cite{nefkens1}. This 
is consistent with the expectation that 
CSV effects are caused by electromagnetic interaction and the
small mass difference between the $u$ and $d$ quarks~\cite{henley1}.

It has been generally assumed that the parton distributions in hadrons
obey charge symmetry. This assumption enables one to relate the parton
distributions in the proton and neutron in a simple fashion,
$u_p(x) = d_n(x), d_p(x) = u_n(x)$, etc. Indeed, charge symmetry is usually 
assumed in the analysis of DIS and DY experiments, which often 
use nuclear targets containing both protons and neutrons. Charge
symmetry is also implicit in the derivation of many QCD sum rules,
including the Gottfried sum rule, the Adler sum rule, and the Bjorken sum rule.

The possibility that charge symmetry could be significantly violated at 
the parton level has been discussed recently by several 
authors~\cite{ma1,ma2,sather,londergan1,rodionov,benesh1,benesh2}. Ma and 
collaborators~\cite{ma1,ma2} pointed out that 
the violation of the GSR 
can be caused by CSV as well as by flavor asymmetry of the nucleon sea.
They also showed that DY experiments, such as NA51 and E866, are subject
to both flavor asymmetry and CSV effects. Using the bag
model, Rodionov et al~\cite{rodionov} showed  
that a significant CSV effect of $\sim 5$\% could exist 
for the ``minority valence quarks" [i.e. $d_p(x)$ and $u_n(x)$] at large
$x$ ($x > 0.4$). A model study~\cite{benesh2} 
of CSV for sea quarks shows that the effect is
very small, roughly a factor of 10 less than for valence quarks.
Londergan \& Thomas have recently reviewed the role of CSV for parton 
distributions~\cite{londergan2}.

Evidence for a surprisingly large CSV effect was recently reported by
Boros et al~\cite{boros1,boros2} 
based on an analysis of $F_2$ structure functions
determined from muon and neutrino DIS experiments. A large asymmetry,
$\bar d_n(x) \sim 1.25 \bar u_p(x)$ for $0.008 < x < 0.1$, is apparently
needed to bring the muon and neutrino DIS data into agreement.
How would this finding, if confirmed by further studies, affect the 
E866 analysis of the flavor asymmetry? First, CSV alone could
not account for the E866 data. In fact, an even larger
amount of flavor asymmetry is required to compensate for the possible CSV 
effect~\cite{boros2}.
Second, there has been no indication of CSV for $x > 0.1$. Thus, the
large $\bar d / \bar u$ asymmetry from E866 for $x > 0.1$
is not affected. 

\noindent{3.4 Parton Distributions of Mesons and Hyperons}

Dilepton production using meson or hyperon beams offers a
means of determining parton distributions of these unstable hadrons. 
Many important features of nucleon parton distributions, such as
the flavor structure and the nature of the nonperturbative sea, find their
counterparts in mesons and hyperons. Information about meson and hyperon 
parton structure could provide valuable new insight into 
nucleon parton distributions. Furthermore,
certain aspects of the nucleon structure, such as the strange 
quark content of the nucleon, could be 
probed with kaon beams.

Pion-induced DY cross sections have been measured in several high-statistics 
experiments~\cite{na10a,na10b,na3a,e615}. These data 
form the basis for a global analysis~\cite{smrs} to 
extract the pion structure functions. Although a large amount of DY data 
exists for $\pi^-$ beams, the corresponding data for $\pi^+$ beams
are surprisingly meager. The $\pi^+$ data are crucial for separating the
valence and sea-quark distributions in pions. The lack of high-statistics 
$\pi^+$ DY data is responsible 
for our poor knowledge of the sea-quark
distributions in pions. Future DY experiments using high-energy 
$\pi^+$ beams ($P_{\pi} > 400~$GeV/c) are required to study the sea of the pion.

The advent of the Fermilab Main Injector (FMI) opens the possibility of 
performing DY measurements with intense pion and 
kaon beams.
The relatively low beam momenta are suitable for 
studying parton distributions at large $x$. 
Londergan et al~\cite{londergan1} suggested
that a comparison between $\pi^+$ and $\pi^-$ DY cross sections 
on hydrogen and deuterium targets could test the charge symmetry of the valence
quark distributions in the nucleon. 

Very little data exist for the kaon-induced DY process. The NA3 
collaboration~\cite{na3b}
obtained several hundred DY events with a $K^-$ beam. By comparing the
$K^-$ with the $\pi^-$ DY data, they found evidence that the $\bar u$
distribution in $K^-$ is significantly softer than in the $\pi^-$.
Because this effect is the largest at large $x_1$, the kaon beam at the 
FMI could be used for further studies.

No data exist for hyperon-induced dilepton production. The observation
of a large $\bar d / \bar u$ asymmetry in the proton has 
motivated Alberg et al~\cite{alberg1,alberg2}
to consider the sea-quark distributions in the $\Sigma$. The meson-cloud
model implies a $\bar d / \bar u$ asymmetry in the $\Sigma^+$ even larger than
that of the proton. However, the opposite effect is expected
from SU(3) symmetry.
Although relatively intense
$\Sigma^+$ beams have been produced for recent experiments at Fermilab,
this experiment appears to be very challenging because of
large pion, kaon, and proton contaminations in the beam.

\noindent{3.5 Nuclear Dependence of the Drell-Yan Process}

\begin{figure}
\centerline{      
\psfig{figure=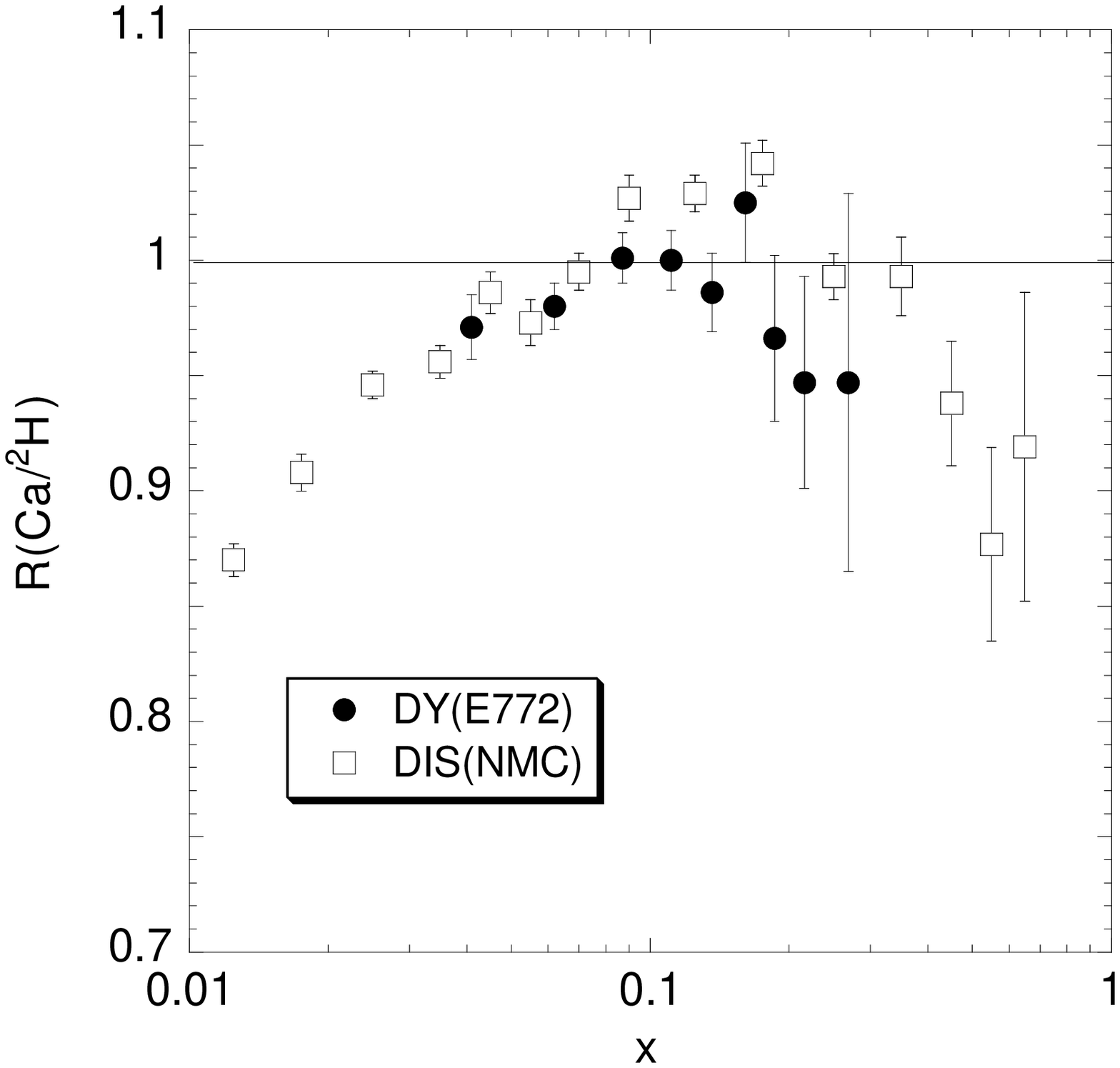,height=4.62in}}
\caption{Comparison of the cross section per nucleon ratios
$\sigma_{Ca}/\sigma_{^2H}$ for DY production~\cite{e772dy} and 
DIS~\cite{nmc}.}
\label{fige772nmc}                                                     
\end{figure}

The famous nuclear dependence of DIS, referred to as the European Muon 
Collaboration (EMC) effect, was discovered in the early 
1980s~\cite{emc,geesaman}. Figure~\ref{fige772nmc} shows the 
general features of the EMC effect,
with data from the NMC~\cite{nmc}.

In 1986, the Fermilab E772 collaboration proposed to measure 
the nuclear dependence of the DY process as a means of 
further elucidating the EMC effect and the issue of parton structure in nuclei 
in general. It is clear from the dominance of the term $u(x_1)\bar 
u(x_2)$ in the kinematic region $x_F\geq 0.2$, that the proton-induced DY 
process provides a view of the nucleon that is complementary to DIS. 
In the E772
configuration, typically more than 90\% of the cross section arises from 
this term. In contrast, only about 15\% of the cross section in DIS at 
$x\sim 0.1$ is due to antiquarks.

\begin{figure}
\centerline{       
\psfig{figure=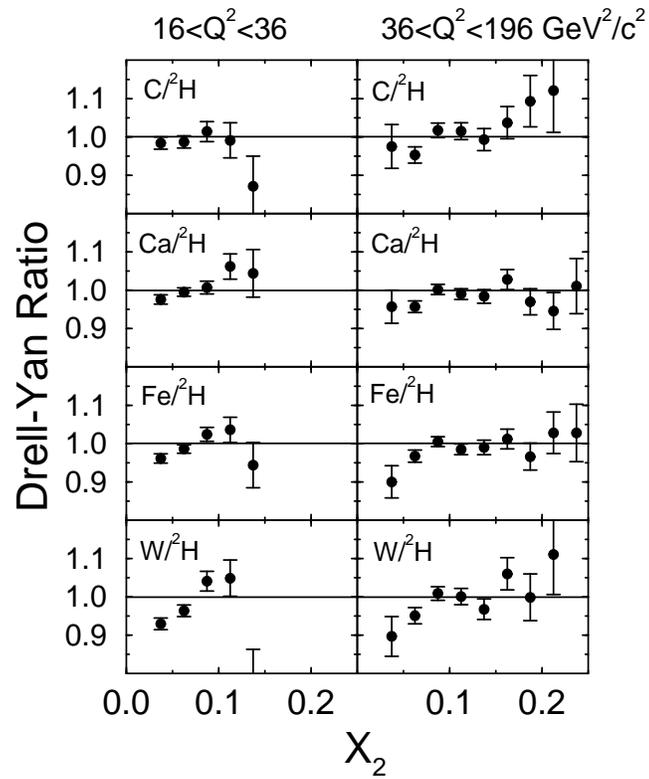,height=4.62in}}
\caption{ Ratio of Drell-Yan cross sections per nucleon for heavy nuclei
to deuterium versus target momentum fraction from E772~\cite{e772dy}. The
two columns show data for different bins of $Q^2=M^2$.}
\label{fige772a}
\end{figure}

The results from E772~\cite{e772dy} (Figures~\ref{fige772nmc} 
and~\ref{fige772a}) show
that there is no enhancement of the antiquark distribution in nuclei. 
Even in the absence of a specific model calculation, the lack of an 
antiquark enhancement in nuclei seems at odds with the picture of nuclei 
as nucleons bound by the exchange of mesons. (Models of the EMC 
effect have been reviewed recently~\cite{geesaman}. We do 
not present a complete overview here.) 

In order to be more quantitative, it is necessary to invoke a model calculation.
The pion excess model, based on the convolution framework 
of Sullivan~\cite{sullivan}, 
is most specific in predicting the nuclear antiquark 
distribution. 
After many years of studying this issue (see e.g.~\cite{ericson}), 
the conventional wisdom is that there 
are ``excess'' pions in nuclei, which are in part 
responsible for nuclear binding~\cite{friman}.
In terms of the pion excess model, the E772 results set stringent limits 
on the collectivity of the $\pi NN$ vertex inside nuclei. It appears 
that the nuclear pion field is not collective at all; there are no more 
pions surrounding the average nucleon in a heavy nucleus than there are in
the weakly bound system, deuterium. This 
contradicts conventional wisdom and is also at odds with 
sophisticated new solutions of the nuclear many-body problem using realistic 
nuclear forces~\cite{friman,pandharipande1,benhar,pandharipande2}. 
When these techniques are applied at central nuclear-matter density, 
they predict 
a substantially increased pion density, $\Delta n_\pi =0.18$ per 
nucleon~\cite{friman,benhar}.\footnote{Reference~\cite{benhar} contains a
correction to the pion excess per nucleon calculated originally 
in Reference~\cite{friman}.}

Using the Sullivan model, Brown et al~\cite{brown}
propose a solution that appears to maintain much 
of the conventional understanding that nuclear binding results from 
meson exchange. They achieve agreement with the E772 antiquark ratio by 
postulating a decrease in the effective masses of hadrons inside nuclei 
resulting from a partial restoration of chiral symmetry. In 
view of the strong recent interest in chiral symmetry restoration 
in relativistic heavy-ion collisions, this is a fascinating
explanation. However, it is far from being universally accepted as the 
answer to the mystery of antiquarks in nuclei. More 
recently, Koltun~\cite{koltun} has advanced another 
suggestion that may reconcile 
the absence of an antiquark enhancement with conventional nuclear theory.

\noindent{3.6 Shadowing}

In the past decade, shadowing $-$ the reduction of the cross section per 
nucleon for nuclear targets at small $x$ $-$ has been very well characterized 
experimentally in DIS~\cite{nmc}. The experimental signature appears 
clearly in Figure~\ref{fige772nmc}, where the DIS ratio falls below unity for 
$x\leq 0.08$. Theoretically, shadowing has been studied 
extensively in the past 10 years (see~\cite{geesaman} for many
seminal references). A recent 
study by Kopeliovich et al~\cite{kopeliovich} presents an alternative 
view to the parton recombination picture of Mueller \& 
Qiu~\cite{mueller}.

Shadowing is also expected in hadronic processes. However, the
beam energies available at fixed-target facilities have limited 
investigations thus far. The reason is clear from the following numerical example. 
In order to have a large enough cross section, one needs $x_1\leq 0.5$. 
This limit, combined with the requirement $M\geq 4$ GeV/c$^2$ at $\sqrt s = 
38.9$ GeV (800 GeV/c fixed target) in Equation~\ref{eq:dykin}, yields $x_2\geq 0.02$.
Similar arguments apply to fixed-target inclusive 
direct photon production. To date, the only 
experimental evidence for shadowing in hadronic reactions is the 
reduction in the nuclear dependence seen in Figures~\ref{fige772nmc} and
\ref{fige772a} from E772~\cite{e772dy}.
The DY and DIS ratios for the same 
targets show similar behavior, although the $x$ range is more limited for 
the DY data.

\begin{figure}                                               
\centerline{                                                 
\psfig{figure=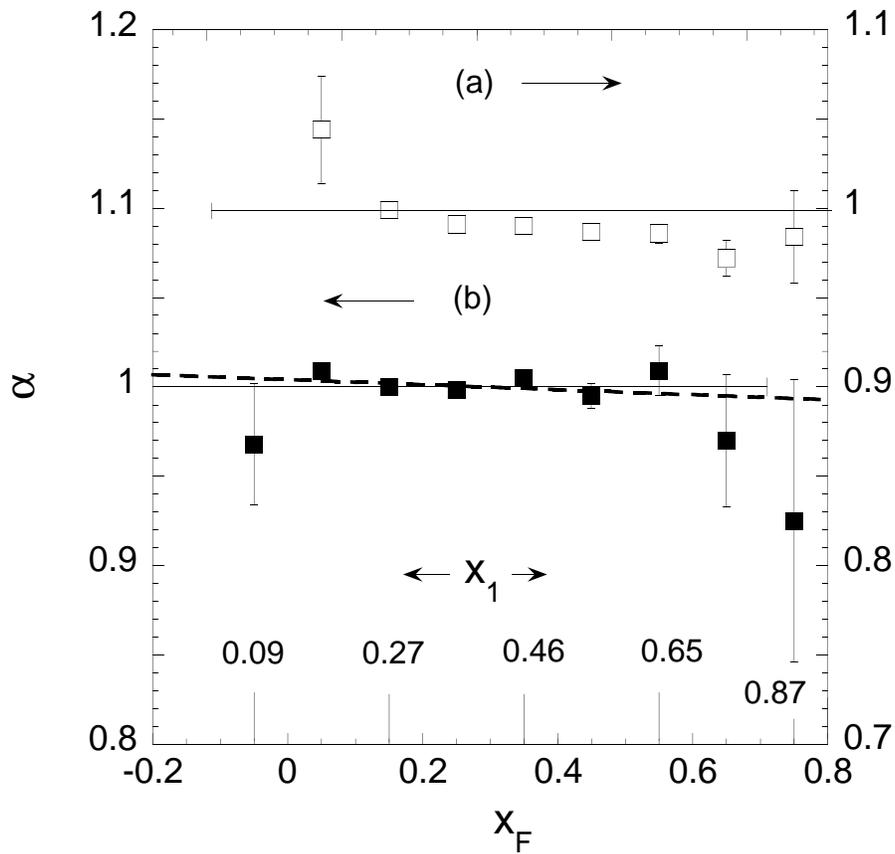,height=4.5in}}
\caption{  Nuclear dependence coefficient $\alpha$ for the Drell-Yan 
process~\cite{e772dy} versus
\xf for (a) $x_2\leq 0.075$, right scale, and  (b) $x_2\geq 0.075$, left 
scale. The thin solid lines show $\alpha =1$. The dashed line is a 
linear least-squares fit to the lower points. Also shown is the 
mean value of $x_1$ for (b) }
\label{fige772b}
\end{figure}

It has been suggested that the reduction in the DY nuclear 
dependence ratio at small $x_2$ is more appropriately ascribed to 
the effects of initial-state 
energy loss~\cite{gavinmilana,quack}. This is unlikely in view of 
Figures~\ref{fige772a} and~\ref{fige772b} and the discussion in 
Section 3.7. Figure~\ref{fige772a} shows 
the nuclear dependence ratios for two different ranges of $Q^2$ ($M^2$). 
It is clear that the reduction at small $x_2$ persists despite the 
signifcant change in the \xf range for the different $Q^2$ cuts.
An alternative view is obtained by fitting all four target ratios 
in terms of the common parameter 
$\alpha$, defined by 
\begin{eqnarray} 
\sigma_A=\sigma_N\times A^{\alpha}. 
\label{eq:alpha} 
\end{eqnarray}
A deviation of $\alpha$ from 
unity indicates a cross section that is not proportional to 
the number of target nucleons. Figure~\ref{fige772b} shows 
$\alpha$ versus \xf for two bins of $x_2$, one in the shadowing 
region({\it upper}) and 
one outside it({\it lower}). The lower plot exhibits no statistically 
significant dependence on $x_F$ or $x_1$,
again showing that the suppression is correlated with the target momentum 
fraction.

\begin{figure}   
\centerline{  
\psfig{figure=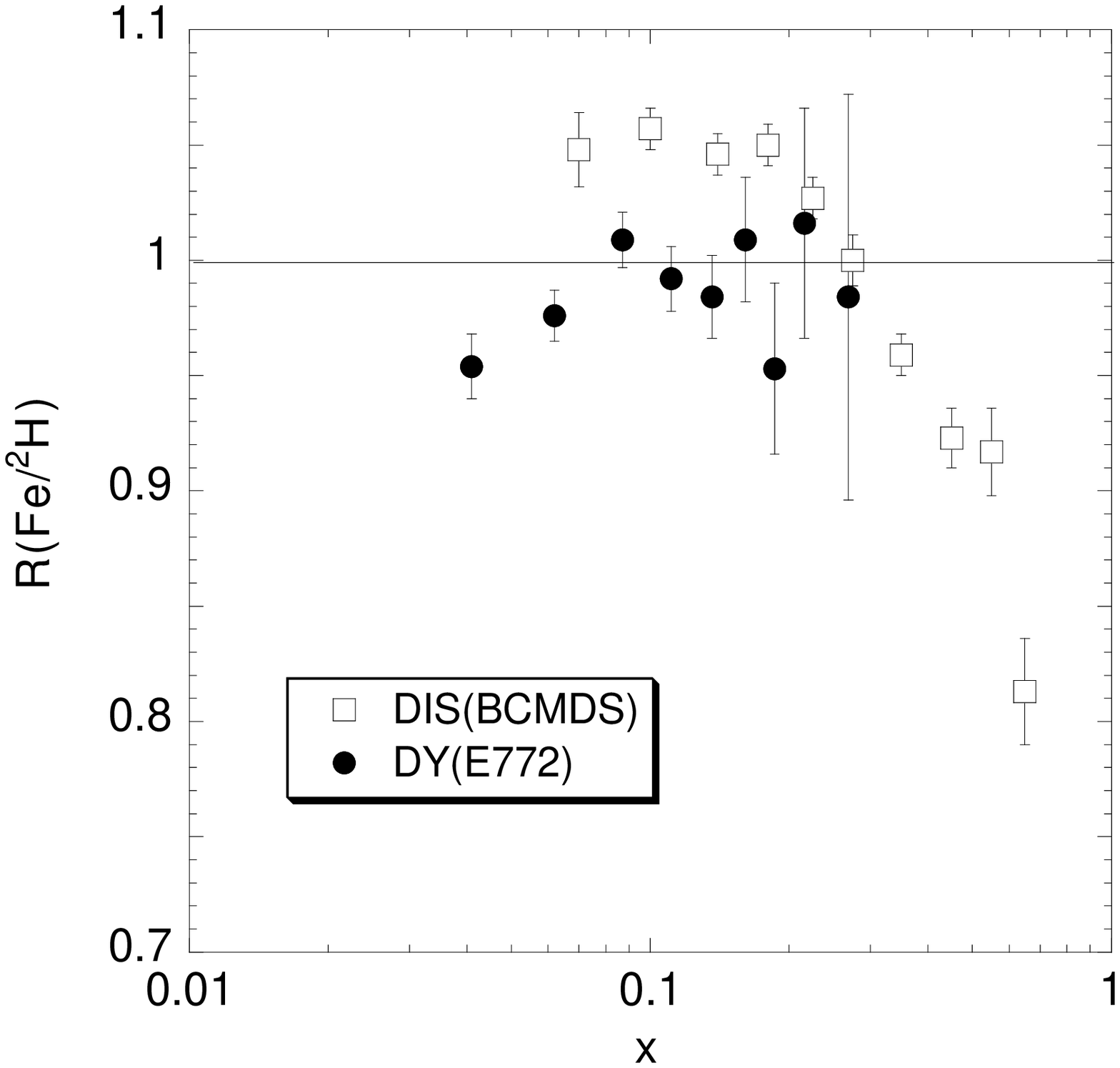,height=4.62in}}
\caption{ Comparison of the cross section per nucleon ratios
$\sigma_{Fe}/\sigma_{^2H}$ for DY production~\cite{e772dy} and
DIS~\cite{bcdms}.}
\label{fige772bcdms}  
\end{figure}

Shadowing in the $p+A$ DY process is largely due to antiquarks in the 
nucleus, unlike in DIS, where quarks dominate for most of the explored 
region. Although shadowing effects are expected for antiquarks and 
gluons, there is no known requirement that they be 
identical~\cite{frankfurt1} to those for quarks. Nevertheless, direct 
comparison of DY and DIS data for Ca versus $^2$H (Figure~\ref{fige772nmc}) 
shows them to be identical within statistics. 
In contrast, in the region $0.1\leq x\leq 0.2$, where DIS shows a modest 
enhancement in the ratio, no corresponding increase occurs in the DY 
ratio. This is also evident in Figure~\ref{fige772bcdms}, which compares
Fe versus $^2$H for DIS and the DY process.

{\noindent 3.7 Multiple Scattering and Energy Loss}

The energy loss of fast partons moving through hadronic matter has 
received extensive theoretical 
attention~\cite{gavinmilana,brodskyhoyer,gyulassywang,luo,bdmps}. 
Recently, Baier et al~\cite{bdmps} have 
derived a connection between the longitudinal energy loss and the mean 
\pt accumulated in multiple parton-nucleon scatterings that is directly 
applicable to experiment. Specifically, 
\begin{eqnarray}
-dE/dz = {1\over 4}\alpha_s N_cp_t^2, \label{eq:dedx}
\end{eqnarray}
where $N_c=3$.
The theory of Baier et al yields the nonintuitive result that the total 
energy loss is proportional to square of the path length traversed.

\begin{figure}
\centerline{
\psfig{figure=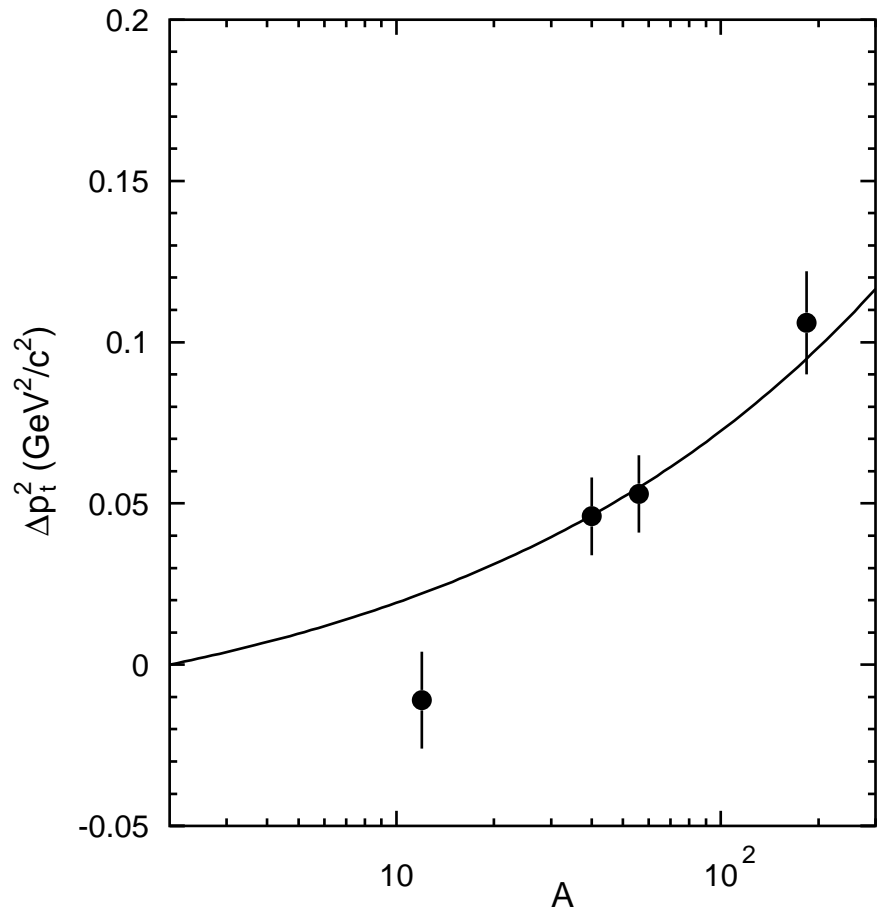,height=5.28in}}
\caption{ $\Delta\langle p_t^2\rangle\equiv
\langle p_t^2\rangle (A)-\langle p_t^2\rangle (^2H)$ versus A for the DY 
process from E772 (\cite{e772ups},PL McGaughey, JM Moss, JC Peng, 
unpublished data).  The solid curve corresponds to
$0.027((A/2)^{1/3} - 1)$.}
\label{fige772dypt2}
\end{figure}

The nuclear dependence of the DY process offers a particularly clean way 
to test this relationship, since multiple scatterings are confined to the 
initial-state quark's traversal of nuclear matter. CERN experiment 
NA10~\cite{bordalo} made the first measurements of the nuclear dependence 
of the DY process as a function of $p_t$. The 
ratio $\sigma_A/\sigma_N$ was found 
to increase with increasing  $p_t$. E772~\cite{e772ups} extended the nuclear 
dependence measurements by determining 
$\Delta\langle p_t^2\rangle$, where $\Delta\langle p_t^2\rangle\equiv
\langle p_t^2\rangle (A)-\langle p_t^2\rangle (^2$H), for DY  production for a 
series of nuclei.
Table 1 lists $\Delta\langle p_t^2\rangle$ from NA10 and E772. 
The E772 values (PL McGaughey, JM Moss, JC Peng, unpublished data)
have changed slightly\footnote{The new evaluation 
is based on an analysis of 
$d\sigma/dp_t^2$ for $^2$H~\cite{e772dimu}, not available previously.} from
those given in Reference~\cite{e772ups}. 
Figure~\ref{fige772dypt2} shows a fit to the E772 data with a 
nuclear dependence of the form 
$\langle p_t^2\rangle (A) \propto A^{1/3}$.

To convert $\Delta\langle p_t^2\rangle$ to an energy loss, one can 
approximate a nucleus by 
a uniform sphere of radius $R_0A^{1/3}$. This yields a mean path length 
of ${3\over 4} R_0A^{1/3}$ for the initial-state partons. Then, using 
Equation~\ref{eq:dedx}, one finds 
$\Delta E\approx 0.59$ GeV for the Tungsten data from E772. In the lab 
frame, $\delta x_1\approx \Delta E/E_{beam} = 7.4\times 10^{-4}$, 
an exceedingly small nuclear effect versus $x_1$ or $x_F$. 
The dashed line in Figure~\ref{fige772b} ({\it lower plot}) is a linear fit
to the data. This slope can be converted into an initial-state energy loss
by using the shape of the accepted spectrum versus $x_1$. This yields 
$\Delta E= 2.0 \pm 1.7$ GeV, which is 
consistent with, but weaker than the limit determined from the \pt 
broadening. 

The absence of a sizable quark energy loss is in 
qualitative accord with the finding of a weak nuclear dependence for the 
production of leading hadrons in DIS~\cite{emchadron,e665hadron}. As has 
been noted before~\cite{bdmps}, however, an interpretation of the 
nuclear dependence of the $p_t$ broadening in photon-induced dijet 
production~\cite{e683} via Equation~\ref{eq:dedx} leads to a very large 
energy-loss effect. We have no new insight into this dilemma. However, it 
is clear that further experimental tests of Equation~\ref{eq:dedx} are 
needed. Understanding how fast partons propagate through hadronic matter 
is crucial for understanding highly relativistic heavy-ion 
collisions and the possible production of the quark-gluon plasma.

{\noindent 4. QUARKONIUM PRODUCTION}

{\noindent 4.1 Quarkonium Production in Hadronic Collisions}

Much of this section focuses on aspects of the nuclear dependence of 
quarkonium production. However, it is important to appreciate the impact 
of the experimental results from the Tevatron Collider on the 
theory of quarkonium production in 
hadronic collisions. Braaten et al~\cite{bfy} have recently 
reviewed this subject. 

In 1995, the CDF collaboration published 
cross sections~\cite{cdfjpsi} for 
vertex-identified (i.e. not arising from $b$-quark decay) \J and \JP production 
at very large $p_t$ --- a particularly advantageous region for comparison 
with theory. The result was that the time-honored color-singlet 
production model underpredicted 
the cross section by factors as large as 30. Motivated by this 
spectacular failure, intense theoretical activity has focused on more general 
production mechanisms, including color-octet channels. 
Color-octet production has become the new paradigm in theoretical 
descriptions of quarkonium production. 

At fixed-target energies, however, the production of quarkonia 
at relatively low \pt presents a challenge to the new 
models~\cite{beneke,tang,schuler}. The effects of color-octet channels 
are reduced, and there are still problems to account for the observed
small polarization. It is 
clear that the continuing theoretical debate 
about mechanisms of quarkonium production will influence, and be 
influenced by, the large nuclear dependence effects discussed below.

{\noindent 4.2 Nuclear Dependence of Quarkonium Production}

In contrast to the DY process, large nuclear effects are found in the 
hadronic production of the \J resonance~\cite{antipov,anderson,corden,badier}.
Whether induced by protons, antiprotons, or pions over a wide range
of beam energies, there is 
a significant reduction in the cross section per nucleon for heavy 
nuclei.\footnote{There are also significant nuclear effects in \J 
production with real~\cite{sokoloff} and virtual photons~\cite{arneodo}. 
These are usually interpreted via the vector-dominance model
in terms of $J/\psi$ absorption on nucleons.}

\begin{figure}   
\centerline{  
\psfig{figure=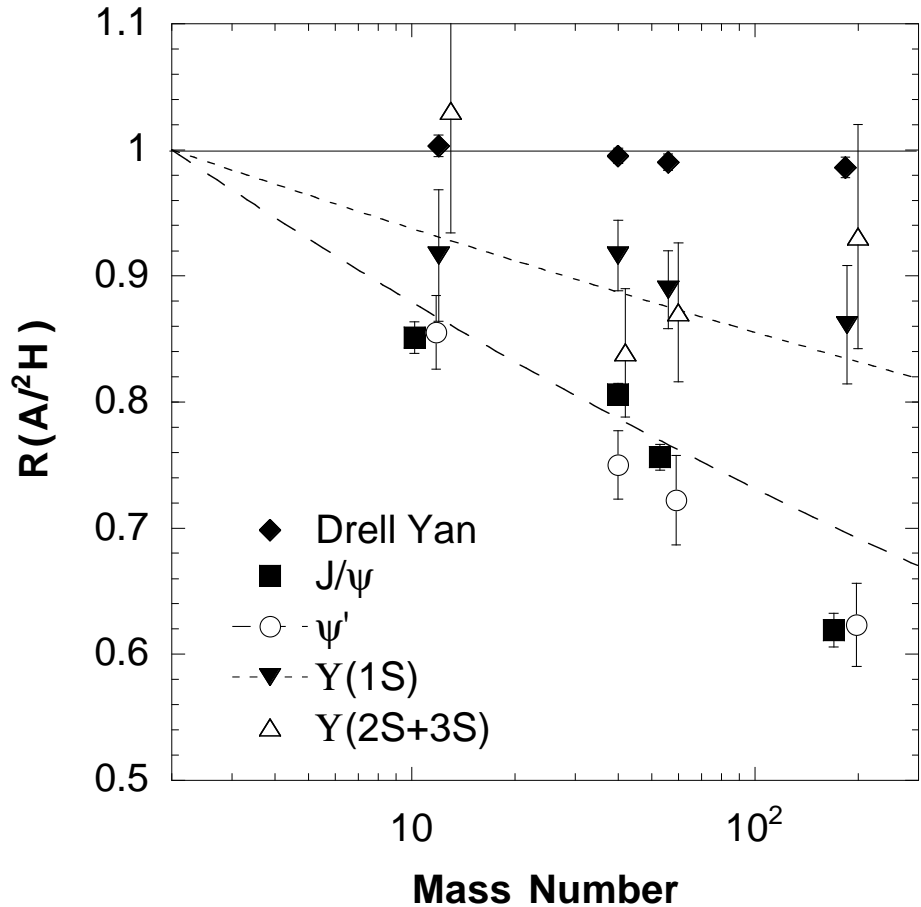,height=4.8in}}
\caption{ Ratios of heavy-nucleus to deuterium
integrated yields per nucleon for 
800 GeV/c proton production of dimuons from the 
Drell-Yan process and from decays of the $J/\psi$, $\psi '$, $\Upsilon 
(1S)$, and $\Upsilon (2S+3S)$ states~\cite{e772dy}. The short-dash and 
long-dash 
curves represent the approximate nuclear dependences for the $b\bar b$ 
and $c\bar c$ states, $A^{0.96}$ and $A^{0.92}$, respectively.}
\label{fige772intQQbar} 
\end{figure}

Using 800-GeV/c protons, E772~\cite{e772jp} was able to resolve the \J and \JP 
resonances for the first 
time in a nuclear-dependence measurement. 
Surprisingly, the E772 collaboration 
found that the nuclear dependences for the two states 
were identical within statistics. Equally unexpected,
they found a 
significant nuclear dependence in the yields of the \u, \up, and \upp 
resonances~\cite{e772ups}, although less than 
that observed for the \J and $\psi'$. Again, within worse 
statistics, the nuclear dependences of the \u and the 
combined \up and \upp were the same.
Figure~\ref{fige772intQQbar} summarizes the integrated nuclear-dependence 
data from E772. 

CERN experiment NA38~\cite{baglin} has confirmed the 
observation of equal nuclear suppression of \J and \JP states in 
$p + A$ collision at 200 GeV/c, employing a modified version of 
the NA10 
spectrometer. NA38 went further to record the first detection 
of charmonium production in relativistic heavy-ion 
collisions~\cite{baglin}. A qualitatively new effect was found:
The ratio $\sigma_{\psi '}/\sigma_{J/\psi}$ decreased by nearly a 
factor of two in $S + U$ collisions with respect to the ratio in 
$p + A$ collisions. The NA38 collaboration and its successor 
NA50~\cite{na50} have developed the capability to measure high-mass 
dimuon events in coincidence with event charge multiplicity and 
calorimetric transverse energy. They have recently recorded interesting 
data on \J and \JP production in $Pb + Pb$ collisions at 157 GeV/c/nucleon (See 
the review by Gerschel \& H\"{u}fner~\cite{gerschel} in this volume.).

\begin{figure}   
\centerline{  
\psfig{figure=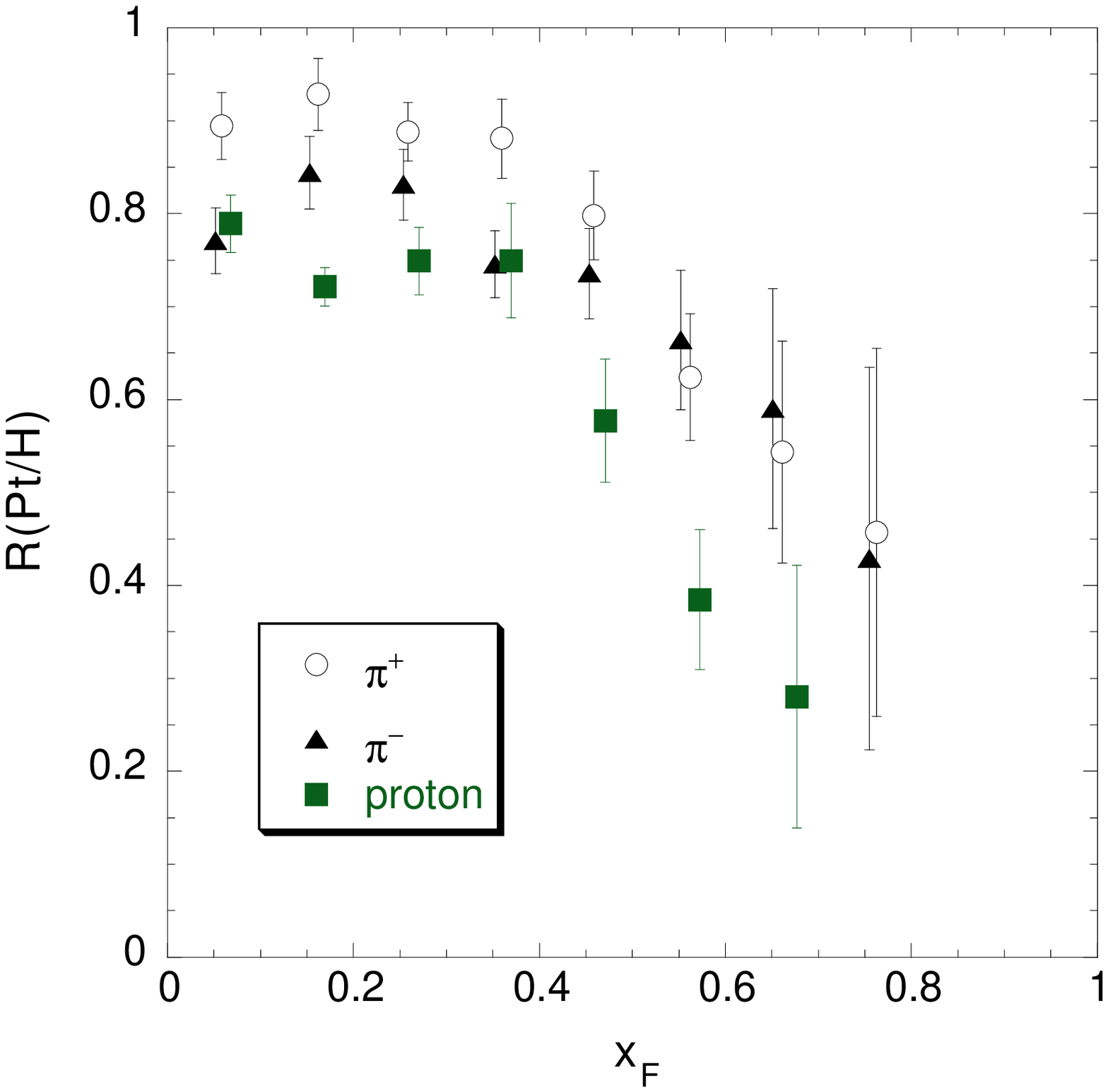,height=4.8in}}
\caption{ Ratios of \J cross sections per nucleon, platinum over
hydrogen, versus \xf for 200 GeV/c $\pi^+$, $\pi^-$, and proton data from
experiment NA3~\cite{badier}.}
\label{figna3jpcross1}                      
\end{figure}

\begin{figure}   
\centerline{  
\psfig{figure=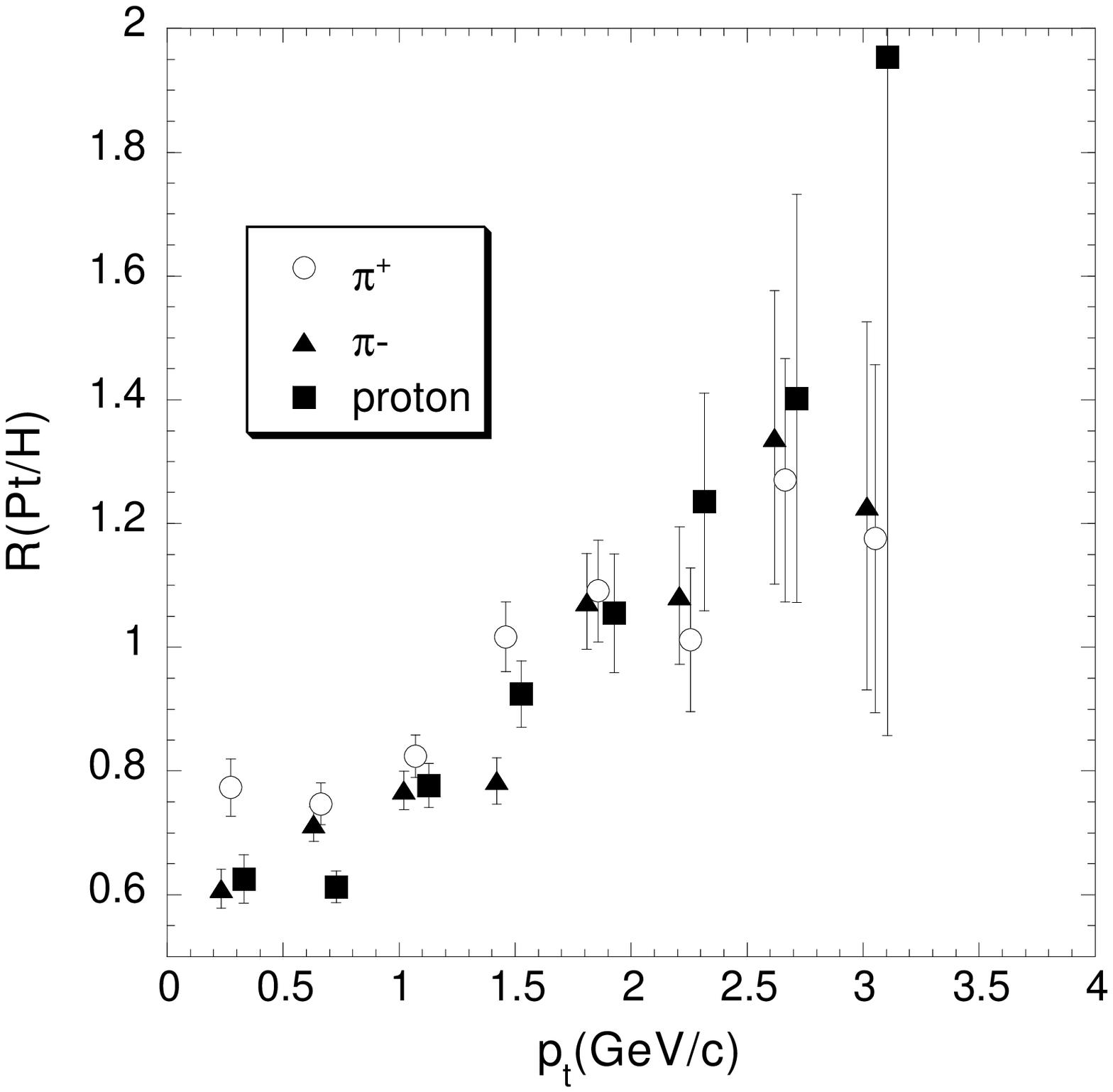,height=4.8in}}
\caption{ Ratios of \J cross sections per nucleon, platinum over
hydrogen, versus \pt for 200 GeV/c $\pi^+$, $\pi^-$, and proton data from
experiment NA3~\cite{badier}.}
\label{figna3jpcross2}                      
\end{figure}

The high-statistics experiment NA3~\cite{badier} 
confirmed observations by earlier experiments of large changes in the nuclear 
dependence as functions of \pt and $x_F$. The NA3 nuclear dependence ratios
versus \xf and \pt have similar shapes for beams of $\pi^-$, $\pi^+$, and 
protons, as shown in Figures~\ref{figna3jpcross1} 
and~\ref{figna3jpcross2}.

\begin{figure}                       
\centerline{                 
\psfig{figure=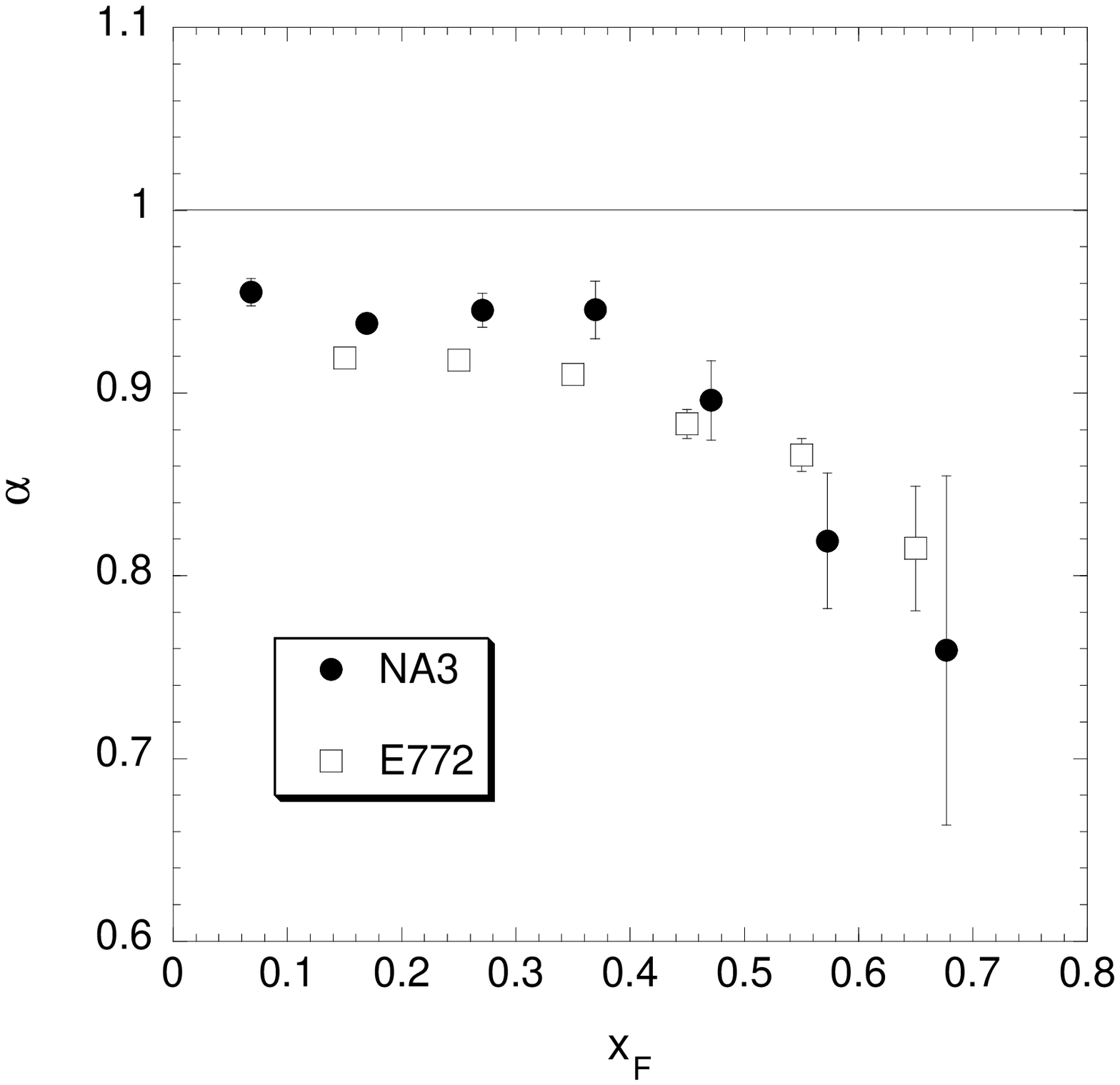,height=4.8in}}
\caption{ $\alpha$ versus \xf for proton-induced \J
production at 200 GeV/c~\cite{badier} and 800 GeV/c~\cite{e772jp}.}
\label{figna3e772xf}  
\end{figure}

\begin{figure}
\centerline{      
\psfig{figure=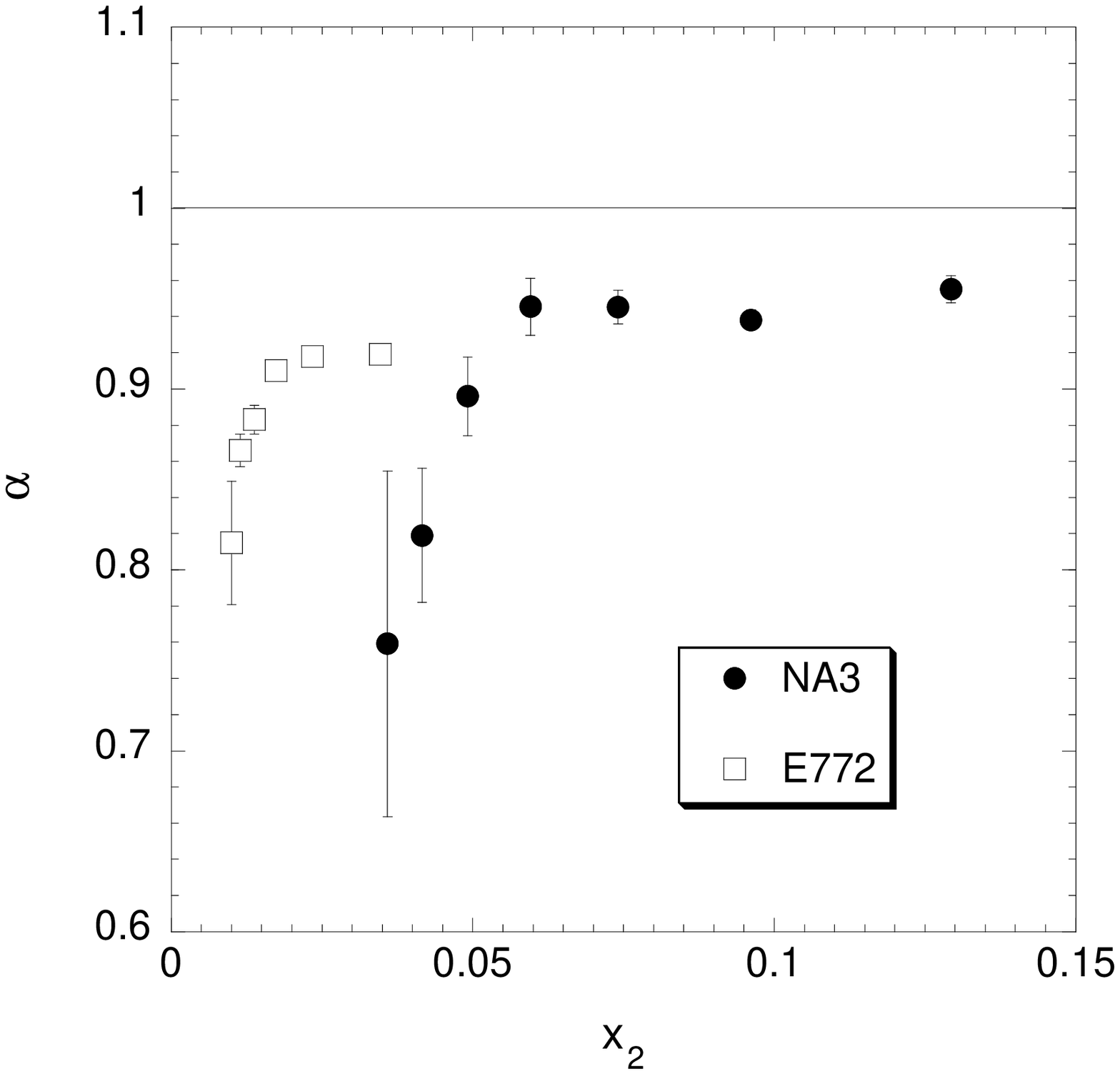,height=4.8in}}
\caption{  $\alpha$ versus $x_2$ for proton-induced \J
production at 200 GeV/c~\cite{badier} and 800 GeV/c~\cite{e772jp}.}
\label{figna3e772x2}
\end{figure}

Figures~\ref{figna3e772xf} and~\ref{figna3e772x2} compare 
proton-induced \J production at 200 GeV/c and 800 GeV/c versus \xf 
and $x_2$. Figure~\ref{figna3e772x2} clearly demonstrates
the lack of scaling with $x_2$. Such scaling would be expected for an 
effect whose origin was primarily due to a nuclear dependence of the 
target structure function $-$ such as shadowing. In addition, 
even though the 800-GeV/c data are in the range where shadowing is 
expected, the observed nuclear dependence is much larger than expected 
for shadowing alone.

The nuclear dependence shows a rise when plotted against \pt for the \J 
(Figure~\ref{figna3jpcross2}) and the \u resonances. Table 2 lists 
$\Delta\langle p_t^2\rangle$ for \J and \u production.
The NA3 values are from 
Badier et al~\cite{badier}. The E772 values (PL McGaughey, JM Moss, JC Peng,
unpublished data)
use a parameterization
that reproduces $d\sigma/
dp_t^2$ for the \u data~\cite{e772dimu} on $^2$H, and the \pt dependence 
of the ratio of cross sections $R(\sigma_A/\sigma_{^2H})$.
Unfortunately, the same analysis could not be performed for the E772's \J 
data owing to the very limited acceptance in $p_t$.

The values of $\Delta\langle p_t^2\rangle$ are similar for the \J and 
the $\Upsilon$, albeit at different beam energies.
The effect is three to four times larger for quarkonia than for DY 
production. A factor of two of this increase may be attributable to 
final-state multiple scattering, which is absent in the DY process.

{\noindent 4.3 Nuclear Dependence of Open Charm Production}

The nuclear dependence of open heavy flavor production is much more 
difficult to measure than that of DY or quarkonium production.
Four experiments employing vertex detectors to identify $D$ 
meson decays have determined 
values of $\alpha$ (see Equation~\ref{eq:alpha}). 
As shown in Table 3,
in contrast to \J and \JP production, there is no evidence of a 
nuclear dependence to open charm production at small $x_F$. No 
measurements have yet been made of the nuclear dependence of open 
beauty production.

{\noindent 4.4 Nuclear Effects in Quarkonium Production and the 
Quark-Gluon Plasma}

Much of the interest in the nuclear dependence of quarkonium production is 
connected to the search for production of the QGP in 
collisions of relativistic heavy ions (see Gerschel \& 
H\"{u}fner~\cite{gerschel}, this volume). In brief, the concept put 
forth by Matsui \& 
Satz~\cite{matsuisatz} is that states such as the \J and \JP would 
not survive if a QGP were formed, since their
radii exceed the effective $c\bar c$ screening length in the QGP. Thus, one 
should observe a suppression of these states in central collisions of heavy 
ions compared with $p + p$ or $p + A$ collisions (where the 
QGP is presumably not formed).
The large nuclear effects found in $p + A$ collisions complicate this 
simple picture. In recent years, much effort has been directed at refining 
the phenomenology of the Matsui-Satz effect in order to identify truly 
new physics that may be manifested in hot-dense matter occupying an extended 
volume~\cite{gerschel}.

To what extent are nuclear effects in quarkonium production
understood? As skeptical experimentalists, we are tempted to 
respond, ``not very well.'' We do not attempt a detailed review of the 
numerous models on the market but rather enumerate 
some of the basic experimental 
elements that must be addressed by models of quarkonium production in 
nuclear systems.

\begin{enumerate}

\item{Significant nuclear effects are found in hadronic production of 
all heavy quarkonium states. Charm is more 
suppressed than beauty, but both exhibit nuclear dependences that are 
much larger than those found for the DY process at comparable values of $Q^2$. 
The few measurements of open 
charm production, based on vertex-identified decays of $D$ mesons,
exhibit no nuclear dependence.}

\item{The nuclear suppression for \J and 
$\psi '$ is the same, within errors, as it is 
for \u and the sum of \up and \upp resonances. Thus, the effect 
does not depend on the final size of the 
hadron. It is important to appreciate that roughly 
half of the total production of the \J originates in decays from 
higher states, including the $\psi '$~\cite{antoniazzi}.}

\item{The nuclear suppression becomes more pronounced at large $x_F$ or 
small $x_2$. However, it does not scale with $x_2$. The nuclear 
dependence versus \xf is similar for production of the \J with 200-GeV/c 
and 800-GeV/c protons.}

\item{A relative nuclear enhancement occurs at large $p_t$. While
this effect is also 
observed for DY production~\cite{e772dy,bordalo}, the rise with 
\pt is much larger for the quarkonium states. The interpretation of the 
\pt dependence in terms of parton energy loss~\cite{bdmps} is more 
complicated than for DY 
production (Section 3.6) because both initial- and final-state effects can 
contribute.}

\end{enumerate}

Although there is now a considerable body of data,
much uncertainty remains in the 
phenomenological description of quarkonium production in nuclei. 
New nuclear-dependence measurements at 
negative values of \xf would rank high as a discriminant between current 
competing models. Such measurements are difficult at fixed-target 
facilities, but it is hoped that they will be carried out in the early years of 
RHIC. 

Hadronic production of quarkonium itself remains a field of intense 
theoretical interest~\cite{bfy}. Many of the desirable new experimental 
investigations relating to the importance of color-octet production, 
e.g. measurements of polarization, would also have an immediate impact 
on issues related to nuclear dependence. Such measurements would also
have a significant impact on the understanding of
quarkonium suppression as a signature of the creation of the QGP.

\noindent{5. FUTURE PROSPECTS}

The next decade will see a qualitative change in experimental capability 
to study dilepton production in hadronic processes. 
Two new hadron colliders, 
RHIC and LHC, will be in operation. Both will have the 
capability of accelerating
and colliding a variety of beams $-$ $p + p$, $p + A$, to $A + A$, where 
$A$ can range to the
heaviest nuclei. RHIC will operate in the range $50\leq\sqrt s\leq 500$ 
GeV/nucleon. Starting in 2000, RHIC will also be able to collide
polarized protons up to $\sqrt s$ = 500 GeV. Although 
specific experimental
capabilities are still being developed at these facilities, 
areas of new opportunity related to the subjects reviewed here can be readily 
identified.

$p + p$ collisions have never been carried out above the energy range
of the CERN intersecting storage rings. Thus, DY production at RHIC and LHC will 
produce new information
for parton density distributions as well as for other issues in QCD that are
accessible via this well-understood reaction. Extension of shadowing studies
in hadronic processes requires the highest 
energy $p + A$ collisions. The DY and
direct photon production are the most readily interpreted processes for such
investigations. The DY process in central collisions of ultra-relativistic 
heavy ions is the most theoretically tractable of the QCD processes in this
exceedingly complex dynamical system. Although DY production
is expected to take place prior to the putative transition to the QGP, there
may be surprises.

$W^{\pm}$ production  in $p + p$ collisions, a first for hadron colliders,
will give a direct indication of the flavor asymmetry of the proton,
not requiring the assumption of charge symmetry~\cite{peng2}. 
Additionally, new tests
of the validity of charge symmetry might be performed via reactions such as
$p+d\rightarrow W^{\pm}$ (\cite{boros2}, S Vigdor, unpublished).
As discussed in Section 3.2, polarized 
proton collisions offer a wealth
of new opportunities for the delineation of 
different aspects of polarized parton structure.
The enormous range of $Q^2$ spanned from 
DY to $W^{\pm}$ and $Z^0$ production will 
give a unique window for observing the interplay 
between quark and gluon spins.

RHIC and the heavy-ion program at the LHC are strongly focused 
on finding evidence 
of the QGP. A major element in this search will be quarkonium suppression.
As we have seen, however, the origins of many 
strong nuclear effects observed in quarkonium production 
in $p + A$ collisions remain mysterious. With the new capabilities 
of dilepton measurement 
in collider detectors, some of these mysteries may
be investigated more easily than in closed-aperture fixed-target experiments.

The new 120-GeV/c Fermilab Main Injector (FMI), scheduled 
to begin operation in 2001, and the proposed 50-GeV/c 
Japanese Hadron Facility~\cite{nagamiya} present capabilities at 
the other end of the $x$ scale. In contrast with the
past 800-GeV/c fixed-target program at Fermilab, which 
could only operate in competition with
the collider, the FMI program will run in parallel. Simple scaling rules show 
that large-$x$
physics is often best carried out at comparatively 
low energies, where cross sections are
largest. As we have seen, measurements of $\bar d/\bar u$ and 
the nuclear dependence of
the antiquark sea at high $x$ require beams from the FMI~\cite{p906}.

In conclusion, dilepton production in hadron-hadron interaction continues
to be a very powerful tool for probing the parton structure of 
hadrons and nuclei,
and for studying the dynamics of the strong interactions. We expect 
significant progress in these areas for the forseeable future, as the
next generation of hadron colliders and high-intensity accelerators begin
operation.

\vskip 0.4in
\noindent{Acknowledgment}

We are grateful to our many collaborators on the E772, E789 and E866
experiments at Fermilab.
In particular, we would like to acknowledge Chuck Brown,
whose expert advice and assistance have been invaluable. This work 
was supported by the US Department of Energy, Nuclear Science 
Division, under contract W-7405-ENG-36.

\newpage

\begin{table}[tbh]
\caption {Values of $\Delta\langle p_t^2\rangle\equiv 
\langle p_t^2\rangle (A)-\langle p_t^2\rangle (^2$H) 
for the DY process. For NA10, the mass range for 
muon pairs was, respectively, $M\geq$ 4.35 GeV/c$^2$ at 140 GeV/c and $M\geq$ 
4.2 GeV/c$^2$ at 286 GeV/c. The mass range of 
the \u (8.5 GeV/c$^2$ $\leq M \leq$ 11 GeV/c$^2$) was also 
excluded. For E772, the range was $M\geq$ 4 GeV/c$^2$, with the 
exclusion of 9 GeV/c$^2$ $\leq M \leq$ 11 GeV/c$^2$.}
\begin{center}
\begin{tabular}{|c|c|c|c|}
\hline
Experiment & Beam &A & $\Delta\langle 
p_t^2\rangle$(GeV$^2$/c$^2)$ \\
\hline\hline
NA10~\cite{bordalo}&140 GeV/c $\pi^-$ & W & $0.16\pm 0.03\pm 0.03$(syst.) \\
\hline
 &286 GeV/c $\pi^-$ & W & $0.15\pm 0.03\pm 0.03$(syst.) \\
\hline
E772~\cite{e772ups}&800 GeV/c p & C & $-0.011\pm 0.015\pm 0.03$(syst.) \\
\hline
 & & Ca & $0.046\pm 0.012\pm 0.03$(syst.) \\
\hline
 & & Fe & $0.053\pm 0.012\pm 0.03$(syst.) \\
\hline
 & & W & $0.106\pm 0.016\pm 0.03$(syst.) \\
\hline
\end{tabular}
\end{center}
\end{table}
\vfill
\eject

\begin{table}[htb]
\caption {Values of $\Delta\langle p_t^2\rangle\equiv
\langle p_t^2\rangle (A)-\langle p_t^2\rangle (^2$H)
for proton-induced production of the \J and \u resonances. The E772 values were 
determined starting with a function, $d\sigma/dp_t^2 = C*(1+(p_t/p_0)^2)^{-6}$, 
with $p_0$ = 2.8 GeV/c, which fits the $^2$H data for the \u 
resonance. The NA3 value for the \J uses $^1$H rather than $^2$H.}
\begin{center}
\begin{tabular}{|c|c|c|c|c|}
\hline
Experiment & State &Beam & A& $\Delta\langle 
p_t^2\rangle$(GeV$^2$/c$^2$) \\
\hline\hline                                         
NA3~\cite{badier}& $J/\psi$ & 200 GeV/c & W & $0.34\pm 0.06$ \\
\hline
E772~\cite{e772ups}& $\Upsilon$ & 800 GeV/c & C & $0.141\pm 0.078\pm 0.03$(syst.) \\
\hline
 & & & Ca & $0.200\pm 0.041\pm 0.03$(syst.) \\
\hline
 & & & Fe & $0.315\pm 0.039\pm 0.03$(syst.) \\
\hline
 & & & W & $0.410\pm 0.078\pm 0.03$(syst.) \\
\hline
\end{tabular}
\end{center}
\end{table}
\vfill
\eject

\begin{table}[htb]
\caption {Nuclear dependence of open charm production via 
vertex-identified $D$ meson decays.}
\begin{center}
\begin{tabular}{|c|c|c|c|}
\hline
Experiment & Beam &$\langle x_F\rangle$ & $\alpha$ \\
\hline\hline
WA82~\cite{wa82}&340 GeV/c $\pi^-$ & 0.24 & $0.92\pm 0.06$ \\ 
\hline
E769~\cite{e769}&250 GeV/c $\pi^-$ & 0.0  & $1.0\pm 0.05\pm 0.02$ \\
\hline
E789~\cite{e789}&800 GeV/c p & 0.031  & $1.02\pm 0.03\pm 0.02$ \\
\hline
WA92~\cite{wa92}&350 GeV/c $\pi^-$ & 0.18 & $0.95\pm 0.06\pm 0.03$ \\
\hline
\end{tabular}
\end{center}
\end{table}

\hspace*{-1.5in}

\hspace*{-0.5in}

\vfill
\eject
\newpage

\vfill
\eject
\newpage

\vfill
\eject
\newpage

\vfill
\eject
\newpage

\vfill
\eject
\newpage

\vfill
\eject
\newpage

\vfill
\eject
\newpage

\vfill
\eject
\newpage
                 
\end{document}